\RequirePackage{silence}
\documentclass[10pt,aps,pra,twocolumn,showpacs,superscriptaddress,nofootinbib]{revtex4-1}
\usepackage[english]{babel} 
\usepackage[usenames, dvipsnames]{color}
\usepackage{graphicx}
\usepackage{bm}
\usepackage{amsmath}
\usepackage{amsthm} 
\usepackage{amssymb} 
\usepackage{times} 
\usepackage{verbatim} 
\usepackage[pdftex,colorlinks=true,urlcolor= blue,linkcolor=Blue,citecolor=RedViolet]{hyperref}

\newcommand{\ket}[1]{|#1\rangle}
\newcommand{\bra}[1]{\langle#1|}

\newcommand{\hy}[1]{\hyperlink{#1}{\color{Gray} #1}}

\usepackage[dvipsnames]{xcolor}

\begin{document}
\title{Gauge Quantum Thermodynamics of Time-local non-Markovian Evolutions}
\author{Fernando Nicacio}
\email{nicacio@if.ufrj.br} 
\affiliation{Instituto de F\'isica, 
             Universidade Federal do Rio de Janeiro, 
             Rio de Janeiro, RJ 21941-972, Brazil}
\author{Raphael N. P. Maia}
\email{pupio@macae.ufrj.br} 
\affiliation{Instituto Polit\'ecnico, 
             Universidade Federal do Rio de Janeiro, 
             Maca\'e, RJ 27930-560, Brazil}                           
\date{\today}

\begin{abstract}
\noindent 
Dealing with a generic time-local non-Markovian master equation, 
we define current and power to be process-dependent as in classical thermodynamics. 
Each process is characterized by a symmetry transformation, 
a gauge of the master equation, and is associated with different 
amounts of heat and/or work. 
Once the symmetry requirement fixes the thermodynamical quantities, 
a consistent gauge interpretation of the laws of thermodynamics emerges.
We also provide the necessary and sufficient conditions for a system 
to have a gauge-independent thermodynamical behavior and 
show that systems satisfying 
Quantum Detailed Balance conditions are gauge-independent. 
Applying the theory to quantum thermal engines, 
we show that gauge transformations can change the machine efficiency,
however, yet constrained by the classical Carnot bound.
\end{abstract}

\maketitle

\section{Introduction}
One of the challenges the nonequilibrium 
quantum thermodynamics is to provide a consistent 
description of the dynamics underlying processes of interaction between 
small or big-scale
quantum systems by means of thermodynamical quantities 
\cite{scovil1959,pusz1978,alicki1979,spohn1978,alicki2018,lindblad1975,
      landi2021,hildred1983,elouard2020,kosloff,colla2022,HOAppl,
      esposito2010,deffner2011,bernardo2020,alipour2016,ahmadi2022,
      vallejo2021,boukobza2006,ribeiro2016,skrzypczyk2014,scopa2018}.
In classical thermodynamics, 
work and heat are process-dependent 
quantities that change the state of the physical system \cite{callen1985}. 
Two states, characterized by their own set of variables 
(pressure, temperature, volume ...), 
can be linked in infinitely many ways, each one differing from the other 
by a chosen process, which can also vary by amounts of 
work and heat extracted/injected in/out the system. 

In a well-established scenario for interacting 
quantum systems 
\cite{BreuerPetr2002,breuer2016,vega2017,Schlosshauer2008,
      lindblad1976,AlickiAppl,maldonado2012,TCLAppl,QDBCfin,
      toscano2022,kalaee2021,davies1974}, 
the dynamics is modeled through a master equation. 
The association of the non-unitary part of this master equation 
with ``heat'' currents dates back to Alicki's paper \cite{alicki1979}, 
leaving work as a manifestation of the unitary piece \cite{pusz1978}. 
However, this dichotomy is not clear as just stated \cite{colla2022,kosloff}. 
As we will show, 
these master equations have a well-known gauge symmetry \cite{BreuerPetr2002}, 
which merges the generators of the unitary part (a Hamiltonian)  
with the ones of the dissipation part (the Lindblad operators),  
while the evolution of the system is kept invariant.  
Since their standard definitions depend on the generators, 
energy, heat currents, and power are not invariant, 
contrary to the system state, which misleads and intertwines 
their notions.
From a logical viewpoint, 
there is no physical reason to withdraw any of these ingredients,
and the only way to deal with the gauge issues is to interpret 
their implications on the physical laws while respecting 
the dynamical invariance. 

In this work, 
we interpret gauge transformations 
as conceivable thermodynamical processes 
keeping the system's evolution unchanged. 
The intertwining of current and power is 
naturally incorporated in our findings 
when we recognize that each process has its distinctive 
fraction of these quantities.
Incorporating the gauge-freedom into thermodynamics, 
the thermodynamical functions turn themselves process-dependent, 
enabling a gauge-consistent definition of a quantum First Law.  

Although gauge-induced contributions will give rise 
to a component of work performed on/by 
the system by/on the reservoir \cite{alipour2016,colla2022}, 
the First Law will consider only two possible ways of energy variation, 
as in classical thermodynamics, throughout heat and work, 
both described by Alicki's formulas \cite{alicki1979} and
their gauge transformations.
We will also explore the existence of systems thermodynamically 
unresponsive to gauges when we write the necessary and sufficient 
conditions for the invariance of the thermodynamical quantities.
Notwithstanding, 
they constitute a small set among all physical systems 
governed by a master equation, 
which enforces the necessity of a gauge-dependent interpretation.  

A general theory for the entropy production for generic 
master equations is still missing \cite{landi2021}, 
which would provide a complete thermodynamical description 
for systems governed by such equations. 
Notwithstanding, 
a recent result \cite{colla2022} 
is used to describe quantum thermal machines 
constituted by a system strongly coupled to thermal reservoirs and, 
interestingly enough, we show that the Carnot bound \cite{callen1985} 
limits the efficiency of these machines for the work and heat provided 
by Alicki definitions and for any of 
their gauge-transformations.

In Section \ref{sec:DynTherm}, 
we review the main hypothesis of our approach, \emph{i.e.}, 
the time-local Master Equation and 
the First Law of quantum thermodynamics. 
Then, we present formal expressions for the gauge transformations 
and, in the sequence, 
we explore the behavior of work and heat rates under gauges. 
Conditions for the invariance of all thermodynamical 
quantities are derived in Section \ref{sec:ThermInv}, 
where we also show that Quantum Detailed Balance is 
sufficient for thermodynamical invariance.
Section \ref{sec:GaugeTherm} provides 
our interpretation of a gauge-dependent thermodynamics, 
where each gauge is associated with a thermodynamical path 
followed by the system and explores the gauge freedom 
for particular processes. 
Remarks about the entropy production 
are in Sec.\ref{sec:EntProd},  
while in Sec.\ref{sec:ExPDM}, 
all presented theory is exemplified within 
the Pure Decoherence Model. 
In Section \ref{sec:atm}, 
quantum thermal machines and the Carnot bound 
are studied under the developed approach.  
As an example, 
we apply all results for a three-level maser. 
We discuss and contextualize our achievements in Sec.\ref{sec:conc} 
with final remarks and outlooks.
%
\section{System Dynamics and Thermodynamics} \label{sec:DynTherm}
The open system scenario, 
see \cite{BreuerPetr2002} for instance, 
considers the evolution of the state of a system $\rm S$, 
initially described by the density operator $\hat \rho_0$, 
as the partial trace 
\begin{equation}\label{eq:dynmap}
\hat \rho(t) =  
{\rm Tr}_\text{E} (\hat U_t \hat \rho_0 \otimes \hat \rho_\text{E} \hat U_t^\dagger)
=:\Phi_t[\hat \rho_0]
\end{equation}
over the Hilbert 
space of another system $\rm E$, where  
the evolution of the joint system ${\rm SE}$ is ruled by 
the unitary operator $\hat U_t$, which is the solution of 
$i\hbar \partial_t\hat U_t = \hat H_{\rm SE}\hat U_t$ 
for the Hamiltonian $\hat H_{\rm SE}$ of the global system. 
This Hamiltonian includes a term relative to the system $\text{S}$, 
another relative to the system E dynamics, 
as well as a coupling energy term. 

When the completely positive trace-preserving (\hypertarget{CPTP}{CPTP}) map 
$\Phi_t$ 
is differentiable with respect to $t$ and invertible, 
\emph{i.e.}, $\Phi_t^{-1}$ exists \cite{maldonado2012}, 
a time-local master equation (\hypertarget{ME}{ME}) \cite{breuer2016,vega2017} 
rules the evolution of the system $\rm S$ (see note \cite{note:ME}):
\begin{equation}\label{eq:mastereq}
\partial_t \hat \rho = 
         \tfrac{i}{\hbar}  [ \hat \rho , \hat H ] 
       - \tfrac{g_{\mu\nu}}{2\hbar}  
             ( 
                    \{    \hat{L}_\mu^\dag \hat{L}_\nu, \hat \rho  \}
                      - 2 \hat{L}_\mu \hat \rho \hat{L}_\nu^\dag   ),
\end{equation}
where all the remaining influence from $\rm E$ are represented by 
Lindblad operators $\hat L_\mu$ ($\mu=1,...,M$) and 
a Hermitian operator $\hat H$, an effective Hamiltonian. 
None of these operators necessarily are pieces of $\hat H_{\rm SE}$, 
since they crucially depend also on $\hat\rho_\text{E}$.
The matrix $\bf g$, whose elements are $g_{\mu\nu}$, 
is a $M \times M$ diagonal matrix defined as 
\begin{equation*}
{\bf g} = {\rm Diag}(\underbrace{1,...,1}_{ m_+  \,\, \text{times} },
                     \underbrace{-1,...,-1}_{m_- \,\, \text{times} }), 
\,\,\, M = m_+ + m_-,
\end{equation*}
and the Einstein convention is employed throughout this text. 
It should be mentioned that ${\bf g}$ may also depend on time%
\footnote{
When all Lindblad operators can be written as 
$\hat L_\mu = \omega_\mu(t) \hat A_\mu$ for 
time-independent $\hat A_\mu$, then 
$g_{\mu\nu} = {\rm sign}(\omega_\mu) \delta_{\mu\nu}$.
Throughout the text, we will denote time-independent Lindblad 
operators by $\hat A_\mu$.},  
that is, $m_\pm$ can change,  
however, this is irrelevant to our results.
Sometimes it will be useful to write Eq.(\ref{eq:mastereq}) 
generically as $\partial_t \hat \rho = \mathfrak{L}(\hat \rho)$
and also to split the generator into a unitary and non-unitary part, 
respectively, defined as the superoperators
\begin{equation}\label{eq:genME}
\begin{aligned}
{\mathfrak U} &:=  \tfrac{i}{\hbar}  [ \bullet , \hat H ],  \\
{\mathfrak D} &:=- \tfrac{g_{\mu\nu}}{2\hbar}  
             ( \{   \hat{L}_\mu^\dag \hat{L}_\nu, \bullet  \}
                   - 2 \hat{L}_\mu \bullet \hat{L}_\nu^\dag   ),
\end{aligned}
\end{equation}   
such that $\mathfrak L = \mathfrak{U} + \mathfrak{D}$.

In principle, we will consider all the operators in (\ref{eq:mastereq}) 
\emph{time-dependent}, 
where memory effects may not be negligible, 
and the non-Markovianity can take place 
despite the absence of a memory kernel \cite{breuer2016,BreuerPetr2002,vega2017}. 
The strict Markovian Lindblad \hy{ME}, 
the one which generates a quantum dynamical semigroup \cite{lindblad1976}, 
will be treated as the particular instance of an autonomous \hy{ME}%
\footnote{%
Autonomous means 
$\partial_t \hat H = \partial_t \hat L_\mu = 
\partial_t g_{\mu\nu} = 0$. 
To gain generality, 
an autonomous \hy{ME} may be different from 
a strict Lindblad \hy{ME}, 
since it is possible to have 
$g_{\mu\nu} \ne \delta_{\mu\nu}$ or, 
equivalently, $m_- \ne 0$.}
with $m_- = 0$. 
Between this and that, the evolution of 
a generic \hy{ME} (\ref{eq:mastereq}) may be Markovian 
with time-dependent operators \cite{BreuerPetr2002,breuer2016,vega2017}.
However, it is only possible to ensure complete positivity
for the dynamics for all times if $m_- = 0, \forall t$.  
For any other $m_-$ values, 
complete positivity may or may not be preserved 
depending on the case \cite{vega2017,breuer2016}.  

As described in \cite{vega2017}, 
the time-local \hy{ME} (\ref{eq:mastereq}) 
is obtained analytically from the CPTP map (\ref{eq:dynmap}).
Surprisingly enough, 
there is no approximations at all,  
nor {\it a priori} assumptions about the state (or the nature) 
of subsystem $\rm E$, 
see \cite{note:ME} for further comments.
Examples of non-Markovian \hy{ME} are \cite{TCLAppl, maldonado2012}, 
some of them highlight the utility of Eq.(\ref{eq:mastereq}) 
in dealing with a strong coupling between $\text{S}$ and $\text{E}$;  
see also the examples and references contained in Refs. 
\cite{breuer2016,vega2017,Schlosshauer2008}.

The operators $\hat H_{\rm SE}$ and $\hat \rho_\text{E}$ determine 
the evolution of $\hat \rho$, however, 
they do not uniquely fix the operators in the \hy{ME}, 
which is manifested by the gauge transformations \cite{BreuerPetr2002}: 
\begin{equation}\label{eq:GaugeTrans}
\begin{aligned}                                                                     
\hat L_\mu & \longrightarrow \hat L_\mu' = U_{\!\mu\nu} \hat L_\nu + \gamma_\mu 
\,\,\,\,\,\, (\mu = 1,...,M),   \\
\hat H & \longrightarrow  
\hat H'= \hat H + \delta\! \hat H, 
\end{aligned}
\end{equation}
with $\nu = 1,...,M$, 
$\gamma_\mu \!:\! t \!\in\! \mathbb R \mapsto \gamma_\mu(t) \in \!\mathbb C$,    
and $U_{\!\mu\nu}\!: t \in \mathbb R \mapsto U_{\!\mu\nu}(t) \in \mathbb C$
are the matrix elements of the pseudo-unitary $M\times M$ matrix $\bf U$ 
satisfying ${\bf U}^\dag {\bf g} {\bf U} = {\bf g}$. 
While each Lindblad operator undergoes an 
affine transformation in (\ref{eq:GaugeTrans}), 
the Hamiltonian is translated by the Hermitian operator
\begin{equation}\label{eq:deltaH}
\delta\! \hat H :=    
          \tfrac{ g_{\mu\nu} }{2i} 
          (\gamma_\mu^\ast U_{\nu\kappa}\hat L_\kappa - 
           \gamma_\mu      U_{\nu\kappa}^\ast\hat L_\kappa^\dagger) + \phi,            
\end{equation}
keeping the \hy{ME} in (\ref{eq:mastereq}) unchanged; 
in above $\phi: t \in \mathbb R \mapsto \phi(t) \in \mathbb R$.   
In fact, the action of gauge (\ref{eq:GaugeTrans}) upon (\ref{eq:genME}), 
after simple calculations, is 
\begin{equation}\label{eq:gGenME}
\begin{aligned}
{\mathfrak U}  \longrightarrow 
{\mathfrak U}'  &= 
\tfrac{i}{\hbar}  [ \bullet , \hat H'] = 
{\mathfrak U}(\hat \rho) + \tfrac{i}{\hbar}[ \bullet , \delta\!\hat H], \\
{\mathfrak D}  \longrightarrow  
{\mathfrak D}' &= 
- \tfrac{g_{\mu\nu}}{2\hbar}  
( \{   \hat{L}_\mu^{\prime\dag} \hat{L}_\nu^{\prime}, \bullet  \}
 - 2 \hat{L}_\mu^{\prime} \bullet \hat{L}_\nu^{\prime\dag}   ) \\
&= {\mathfrak D} - 
\tfrac{i}{\hbar} [ \bullet , \delta\!\hat H], 
\end{aligned}
\end{equation}
which express the invariance of the global generator:  
${\mathfrak{L}'} := 
{\mathfrak U}' + {\mathfrak D}' = {\mathfrak U} + {\mathfrak D} =
{\mathfrak{L}}$.

The set of all transformations in (\ref{eq:GaugeTrans}) 
is the symmetry group of the Lindblad equation,
defined by the product rule \cite{colla2022} 
$(\gamma_\mu'', {\bf U}'', \phi'' )
 (\gamma_\mu',  {\bf U}' , \phi'  )$ 
$ = (\gamma_\mu, {\bf U}, \phi)$, 
where
\begin{equation*}
\begin{aligned}
&\gamma_\mu = \gamma_\mu'' +  {U}_{\mu\nu}'' \gamma_\nu', 
\quad {\bf U} = {\bf U}'' {\bf U}' \\  
&\phi = \phi' + \phi'' + 
{\rm Im}(\gamma_\mu^{\prime\prime\ast} g_{\mu\nu} U_{\nu\kappa}''\gamma_\kappa').  
\end{aligned}
\end{equation*}
These symmetries are 
inherently associated with the origin of a \hy{ME},  
which is only justified when, whatever the reason, 
there is a lack of knowledge or control 
about the system and its surroundings, 
this lack manifests as the gauges. 
In other words, 
if one has complete control over the global system 
($\hat H_{\rm SE}$, $\hat \rho_\text{E}$, and $\hat \rho_0$),  
the dynamics will be promptly described by the global 
unitary evolution, and the gauges are as useless as a \hy{ME}. 
 
When an experimentalist does not know the microscopic details about 
the whole system-environment state, 
he/she is left with standard tomographic procedures  
for measurements of the system state or 
of the generator $\mathfrak{L}$, both gauge-invariant. 
The obtained generator will fit with any operators 
connected by a gauge transformation (\ref{eq:GaugeTrans}). 
See chapter 8 of the book \cite{nielsen2001} 
for a description of 
tomography for quantum state and processes 
and its experimental references.

The dichotomy heat-work naturally emerges \cite{alicki1979} 
for the \hy{ME} (\ref{eq:mastereq}) 
if the internal energy of the system is taken as the mean value 
$\langle \hat H \rangle := {\rm Tr}(\hat H \hat \rho)$, 
once a temporal derivative reads
\begin{equation} \label{eq:firstlaw}
\partial_t \langle \hat H \rangle = \mathcal J + \mathcal P, 
\end{equation}
with the \emph{Lindblad total current} and the \emph{Lindblad power}
defined, respectively, as 
\begin{equation}\label{eq:powcurr}
\mathcal J := {\rm Tr}( \hat H \partial_t \hat \rho  ) = 
              {\rm Tr}[ \hat H {\mathfrak D}(\hat \rho) ], \,\,\,\, 
\,\,\, 
\mathcal P := \langle \partial_t \hat H\rangle.   
\end{equation}
The expression for the total current above 
is a consequence of ${\rm Tr}[\hat H {\mathfrak U}(\hat \rho)] = 0$, 
which shows that there is no contribution 
of the unitary part and that the current is
proper to the interaction with subsystem $\rm E$.  
Writing  $\mathfrak D$ explicitly as in (\ref{eq:genME}), 
\begin{equation} \label{eq:sumcurr}
{\mathcal J} = \sum_{\mu=1}^M {\mathcal J}_\mu = 
-  \tfrac{1}{\hbar} g_{\mu\nu} \, {\rm Re} \,
\langle \hat{L}_\mu^\dagger [ \hat{L}_\nu , \hat{H} ]  \rangle,
\end{equation}
{\it i.e.}, 
the total current is the sum of the currents 
due to each Lindblad operator
\begin{equation}\label{eq:LindCurr}
{\mathcal J}_\mu :=  
-  \tfrac{1}{\hbar}{\delta_{\mu\kappa}\,g_{\kappa\nu}} \, {\rm Re} \,
\langle \hat{L}_\kappa^\dagger [ \hat{L}_\nu , \hat{H} ] \rangle.  
\end{equation}

The above formulas, established by Alicki in \cite{alicki1979},  
constitute the standard definitions for the thermodynamical quantities 
associated with the \hy{ME} \cite{Note:AlickiDef}.
However, the invariance of the \hy{ME} under any gauge (\ref{eq:GaugeTrans})
does not apply to the thermodynamical functions \cite{kosloff}, 
which is the starting point of our work.

\subsection{Energy Non-Invariance}
The first step towards understanding the relation between gauges 
and thermodynamics is to consider the energy of the system. 
Applying the gauge transformation (\ref{eq:GaugeTrans}) in the mean energy,
$\langle\hat H \rangle\longrightarrow \langle\hat H'\rangle = 
\langle\hat H \rangle + \langle\delta\!\hat H\rangle$,
taking the temporal derivative and using \eqref{eq:firstlaw}, 
one obtains
\begin{equation}\label{eq:gFirstLaw}
\begin{aligned}
\partial_t \langle \hat H' \rangle 
&=  
       \partial_t \langle \hat H \rangle + 
       \partial_t \langle \delta \! \hat H \rangle  =                     
 \mathcal J + \mathcal P + \partial_t \langle \delta \! \hat H \rangle.                
\end{aligned}
\end{equation}
We can thus already conclude that the mean energy of the system, 
as well as its variation, changes for an applied gauge.
In the sequence, 
we perform the remaining derivative in (\ref{eq:gFirstLaw}):
\begin{equation}\label{eq:derivdeltaH}
\begin{aligned} 
\partial_t \langle \delta\! \hat H \rangle &= 
{\rm Tr}(\delta\! \hat H \, \partial_t \hat \rho  ) +  
\langle \partial_t \delta\! \hat H \rangle                \\ 
&={\mathcal J}_{\! \delta \! \hat H} 
+ {\mathcal C}_{\! \delta \! \hat H} 
+ \langle \partial_t \delta\! \hat H \rangle,  
\end{aligned}
\end{equation}
where the last equality is obtained by inserting 
the \hy{ME} \eqref{eq:mastereq}, using (\ref{eq:genME}), 
and defining 
\begin{equation} \label{eq:curpowdeltaH}
\begin{aligned} 
{\mathcal J}_{\! \delta \! \hat H} &:= 
{\rm Tr}[ \delta\! \hat H {\mathfrak D}(\hat \rho) ]=
- \tfrac{g_{\mu\nu}}{\hbar}\,
{\rm Re}\langle \hat{L}_\mu^\dagger \, [ \hat{L}_\nu , \delta\! \hat H ] 
\rangle, \\
{\mathcal C}_{\! \delta \! \hat H} &:=  
{\rm Tr}[ \delta\! \hat H {\mathfrak U}(\hat \rho) ] = 
 \tfrac{i}{\hbar} \langle [ \hat H, \delta\! \hat H]\rangle.  
\end{aligned}
\end{equation}
Finally, collecting all these calculations, 
\begin{equation}\label{eq:gFirstLaw2}
\partial_t \langle \hat H' \rangle = 
\mathcal J + \mathcal P + 
{\mathcal J}_{\! \delta \! \hat H} +
{\mathcal C}_{\! \delta \! \hat H} +
\langle \partial_t \delta\! \hat H \rangle,                
\end{equation}
which is the action of 
the gauge transformations into Eq.(\ref{eq:firstlaw}).

While the dynamics governed by the \hy{ME} is invariant, 
the quantum counterpart of the First Law ---  provided by 
Alicki's definitions in Eq.(\ref{eq:firstlaw}) --- 
is not invariant and acquires other components. 
Fortunately, 
all gauge-induced contributions in \eqref{eq:gFirstLaw2} 
have a precise physical meaning.  
Compared with $\mathcal P$ in (\ref{eq:powcurr}), 
$\langle \partial_t \delta\! \hat H \rangle$ is a power component 
due to the Hamiltonian $\delta\! \hat H$, 
as well as, 
${\mathcal J}_{\! \delta \! \hat H}$ in (\ref{eq:curpowdeltaH}) 
is a current with the same structure as $\mathcal J$ in (\ref{eq:sumcurr}). 
In Eq.(\ref{eq:curpowdeltaH}), 
${\mathcal C}_{\! \delta \! \hat H}$ is the mean value of 
the unitary evolution (the commutator with the system Hamiltonian $\hat H$) 
of the operator $\delta \! \hat H$, 
thus another power component. 
Noteworthy, 
all these gauge contributions are mean values of 
Hermitian operators (observables), 
as well as ${\mathcal J}$ and ${\mathcal P}$. 

\subsection{Covariance of the First Law}
To go deeper into our analysis, 
the individual behavior of each thermodynamical quantity in Eq.(\ref{eq:firstlaw}) 
will be explored in order to give a precise meaning to formula (\ref{eq:gFirstLaw2}). 

Each of the system currents (\ref{eq:LindCurr}) 
under a gauge (\ref{eq:GaugeTrans}) becomes 
\begin{eqnarray}\label{eq:gLindCurr}
{\mathcal J}_\mu \longrightarrow {\mathcal J}'_\mu = 
%
&-& \tfrac{1}{\hbar} \delta_{\mu\kappa} g_{\kappa\nu} \, {\rm Re} 
\langle \hat{L}_\kappa'^\dagger [ \hat{L}'_\nu , \hat{H} +\delta \! \hat H ] \rangle,                                                                          
\end{eqnarray}
where $\hat L_\mu'$ is defined in 
(\ref{eq:GaugeTrans}) and $\delta\! \hat H$ in (\ref{eq:deltaH}). 
Summing up for all $\mu$, 
after tedious but straightforward manipulations,
the transformed current reads
\begin{eqnarray} \label{eq:gCur}
{\mathcal J} \longrightarrow 
{\mathcal J}'&:=& \sum_{\mu=1}^M {\mathcal J}'_\mu             
                 = {\mathcal J}   
                 + {\mathcal J}_{\! \delta \! \hat H}
                 + {\mathcal C}_{\! \delta \! \hat H},   
\end{eqnarray}
{\it i.e.}, 
the total current, which itself is non-invariant, 
accounts for both quantities (\ref{eq:curpowdeltaH}), 
see Eq.(\ref{eq:derivdeltaH}).  
For the Lindblad power $\mathcal P$ in (\ref{eq:powcurr}), 
applying the symmetry transformation (\ref{eq:GaugeTrans}), one gets
\begin{equation} \label{eq:gPower}
\mathcal P \longrightarrow \mathcal P' = 
\langle\partial_t\hat H' \rangle = 
\langle\partial_t (\hat H + \delta\! \hat H ) \rangle=
\mathcal P  + 
\langle \partial_t \delta\! \hat H \rangle
\end{equation}
and the remaining gauge contribution 
$\langle \partial_t \delta\! \hat H \rangle$ in (\ref{eq:derivdeltaH}), 
comes from the non-invariance of the power.

Consistently, 
it should be noted that $\mathcal J'$ in (\ref{eq:gCur}) is equal to
\begin{equation}\label{eq:Transfcur2}
\mathcal J' = {\rm Tr}( \hat H' \partial_t \hat \rho  ) 
= {\rm Tr}[ \hat H {\mathfrak D}(\hat \rho) ]
+ {\rm Tr}( \delta\!\hat H \partial_t \hat \rho ),  
\end{equation}
in accordance with $\mathcal J$ in Eq.(\ref{eq:powcurr}) 
and ${\rm Tr}(\delta\! \hat H \, \partial_t \hat \rho  ) 
={\mathcal J}_{\! \delta \! \hat H} + {\mathcal C}_{\! \delta \! \hat H}$
from (\ref{eq:derivdeltaH}). 

Summing the above expression for $\mathcal J'$ 
with $\mathcal P'$ in (\ref{eq:gPower}),  
one attains a gauge \emph{covariant} expression for the First Law: 
\begin{equation}\label{eq:TransfmeanH}
\partial_t \langle \hat H' \rangle = 
{\rm Tr}( \hat H' \partial_t \hat\rho + \hat\rho \, \partial_t \hat H' ) 
= \mathcal J' + \mathcal P',    
\end{equation}
which is simply a consequence of the linearity 
of both temporal derivative and trace operations  
in Eqs.(\ref{eq:firstlaw}) and (\ref{eq:powcurr}). 

The non-invariance of the mean energy 
$\langle\hat H'\rangle\ne \langle\hat H\rangle$ 
is the root of all gauge-induced contributions, 
see Eqs.(\ref{eq:gFirstLaw}) and (\ref{eq:derivdeltaH}), 
and deserves some comments regarding the energy-conservation. 
Gauges are ascribed to information lack 
owing to the trace in (\ref{eq:dynmap}) \cite{BreuerPetr2002},
which means that 
many operators $\hat H_\text{SE}$ and $\hat \rho_\text{E}$ 
raise the same \hy{ME} (\ref{eq:mastereq}), 
which only contains information about the system state $\hat \rho$. 
The non-invariance of the energy, 
$\langle\hat H'\rangle = 
\langle\hat H \rangle + \langle\delta\!\hat H\rangle$,
is thus a comparison between two different global systems $\text{SE}$. 
For each of these, 
the mean value of the global Hamiltonian, 
say $\hat H_\text{SE}$ or $\hat H'_\text{SE}$, 
determines the total energy, 
and the partial trace in (\ref{eq:dynmap}) 
selects the corresponding part of the system, determined by 
$\hat H$ or $\hat H'$, 
which does not contradict the global energy conservation 
for each global system.

\subsection{The Interplay among Generators}\label{sec:TIAG}
Ultimately, 
a gauge transformation is an interplay 
of the operator ${\delta\! \hat H}$ between the generators 
$\mathfrak{U}$ and $\mathfrak{D}$, see Eq.(\ref{eq:gGenME}). 
To explore this observation thermodynamically, 
we will make a digression and consider a 
slightly different situation without gauge transformations. 

Consider a \hy{ME} given by
\begin{equation}\label{eq:intGen}
\partial_t \hat \rho = {\mathfrak L}_1(\hat \rho) :=  
\mathfrak{U}'(\hat \rho) + \mathfrak{D}(\hat \rho), \,\,\, 
\mathfrak{U}' = \mathfrak{U} + 
\tfrac{i}{\hbar} [ \bullet , \hat V], 
\end{equation}
with $\mathfrak{U}$ and $\mathfrak D$ as in (\ref{eq:genME}), 
{\it i.e.}, the dynamics is described by a \hy{ME} (\ref{eq:mastereq}) 
with the Hamiltonian $\hat H' = \hat H + \hat V$ 
and Lindblad operators $\hat L_\mu$.
Equivalently, it is possible to write for the same \hy{ME} that
\begin{equation}\label{eq:intGen2}
\partial_t \hat \rho = {\mathfrak L}_2(\hat \rho) 
:= \mathfrak{U}(\hat \rho)  + \mathfrak{D}''(\hat \rho) , 
\,\,\,
\mathfrak{D}'' := {\mathfrak D} + 
\tfrac{i}{\hbar} [ \bullet , \hat V].  
\end{equation}

If and only if the potential term $\hat V$ 
is a Hermitian linear combination of $\hat L_\mu$,  
the same Lindblad operators appearing in the generator ${\mathfrak D}$, 
it is possible to write $\hat V = \delta \!\hat H$ 
for some set of functions $\{U_{\mu\nu},\gamma_\mu, \phi\}$ 
in (\ref{eq:deltaH}). 
As learned in (\ref{eq:gGenME}), 
for $\hat V$ written as (\ref{eq:deltaH}), 
it is possible to regroup the generator 
$\mathfrak{D}''$ in (\ref{eq:intGen2}) as 
\[
\mathfrak{D}''(\hat \rho) = 
- \tfrac{g_{\mu\nu}}{2\hbar}  
( \{ \hat{L}_\mu^{\prime\prime\dag} \hat{L}_\nu^{\prime\prime}, \hat \rho  \}
- 2 \hat{L}_\mu^{\prime\prime} \hat \rho \hat{L}_\nu^{\prime\prime\dag} )
\]
with $\hat L''_\mu = U_{\mu\nu}\hat L_\mu - \gamma_\mu$, 
note the different signs in \eqref{eq:intGen2} and in \eqref{eq:gGenME}.  
Therefore, the same evolution is ruled by an equivalent \hy{ME}, 
$\partial_t \hat \rho = {\mathfrak L}_2(\hat \rho)$, with 
$\hat L''_\mu$ and $\hat H = \hat H'-\hat V$. 

From the thermodynamical point of view, 
energy, currents, and power definitions 
are corrupted by the interplay of $\hat V = \delta\!\hat H$, 
and the reason is the same as before: 
the mean energy of the system changes, as well as, 
its distribution among heat and work. 
For the generator $\mathfrak{L}_1$ in (\ref{eq:intGen}), 
we have ${\rm Tr}[ \hat H' {\mathfrak U}'(\hat \rho) ] = 0$, 
thus
\[
\partial_t \langle \hat H' \rangle = 
{\rm Tr}[ \hat H' {\mathfrak L}_1(\hat \rho) ] + 
\langle \partial_t \hat H' \rangle = 
{\mathcal J} + {\mathcal J}_{\! \delta \! \hat H} + {\mathcal P}' 
\]
for 
${\mathcal J}$ in (\ref{eq:powcurr}), 
${\mathcal J}_{\! \delta \! \hat H}$ in (\ref{eq:curpowdeltaH}), 
and 
$\mathcal P'$ in (\ref{eq:gPower}).
However, the generator $\mathfrak{L}_2$ in (\ref{eq:intGen2}) reads
\[
\partial_t \langle \hat H \rangle = 
{\rm Tr}[ \hat H {\mathfrak L}_2(\hat \rho) ] + 
\langle \partial_t \hat H \rangle = 
{\mathcal J} - {\mathcal C}_{\! \delta \! \hat H} + {\mathcal P}, 
\]
for 
${\mathcal J}$ and $\mathcal P$ both in (\ref{eq:powcurr}), 
and 
${\mathcal C}_{\! \delta \! \hat H}$ in (\ref{eq:curpowdeltaH}). 
Therefore, like for a gauge transformation, 
a simple redistribution of the potential $\hat V$ among 
the \hy{ME} generators breaks the dichotomy heat-work 
given by Alicki's definitions in (\ref{eq:powcurr}).  
Besides, the gauge-induced terms in (\ref{eq:derivdeltaH}) 
appear on both versions of the First Law above  
as a consequence of the redistribution, 
which indicates that those terms need to 
be correctly accounted when considering 
the thermodynamics of a quantum system governed by a \hy{ME}.

Far from pure theoretical curiosity, 
this interplay may happen in truly physical systems, 
which is the case of the infamous phenomenon 
of resonance fluorescence, 
see for instance \cite{BreuerPetr2002,hildred1983}. 
%
%
In such case, 
the dynamics is modeled by a Markovian Lindblad \hy{ME} (\ref{eq:mastereq})
with $M = 2$, $\mu \in \{ +, -\}$, 
$g_{\mu\nu} = \delta_{\mu\nu}$ and   
\[
\begin{aligned}
\hat H' &= \tfrac{1}{2}\hbar \omega_0\, \hat \sigma_z + \hat V, \,\,\, 
\hat V :=
\hbar \Omega ( \text{e}^{-i \omega t} \hat \sigma_+ 
             + \text{e}^{+i \omega t} \hat \sigma_- ), \\          
\hat L_{\pm} &= \lambda_\pm\, \hat \sigma_\pm, \,\,\, 
\lambda_\pm := \sqrt{\hbar\Gamma(\bar n\mp 1/2+1/2)}, 
\end{aligned}
\]
where $\hat \sigma_z$ and $\hat \sigma_\pm$ 
are the standard {\rm SU}(2) matrices,   
$\omega_0$ is an atomic frequency transition, 
$\omega$ is a coherent laser frequency, 
the Rabi frequency is $\Omega$, 
the decaying rate is $\Gamma$, 
and $\bar n$ is the 
mean-occupation number of a reservoir.
The interaction term can be written as $\hat V = \delta\!\hat H$ 
in Eq.\eqref{eq:deltaH} using the functions
\[
\gamma_\pm = - i\hbar\Omega \lambda_\pm^{-1} \text{e}^{ \pm i\omega t}, \,\,\, 
\phi =0, \,\,\, U_{\mu\nu} = \delta_{\mu\nu}.  
\]
The \hy{ME} can thus be written as in (\ref{eq:intGen}) 
with $\hat H'$ and $\hat L_\mu$ above 
or equivalently by (\ref{eq:intGen2}) for the operators
\[
\hat H = \tfrac{1}{2}\hbar \omega_0\, \hat \sigma_z, \,\,\,  
\hat L''_{\pm} = \lambda_\pm\, \hat \sigma_\pm - \gamma_\pm.  
\]
Remarkably the atom-field interaction 
$\hat V = \hat H' - \hat H = {\delta\! \hat H}$ 
manifests itself as an affine transformation of 
the Lindblad operators, like in Eq.(\ref{eq:GaugeTrans}). 

After all, 
although we start saying that we were not dealing with gauges to treat the interplay, 
it can be understood as such. 
Consider the evolution given by \eqref{eq:intGen} with 
$\hat H = \hat H_0 + \hat V$ and $\hat V$ written as (\ref{eq:deltaH}).   
The gauge transformation $\hat L'_\mu = U_{\mu\nu}\hat L_\nu - \gamma_\mu$ 
translates the Hamiltonian $\hat H$ by $\delta\!\hat H = -\hat V$, 
see Eq.(\ref{eq:GaugeTrans}),   
and the transformed Hamiltonian is $\hat H'= \hat H_0$.  
After the gauge, 
the same dynamics will be governed by a \hy{ME} 
with generators written as (\ref{eq:gGenME}). 

\subsection{Statistical Interpretation of Heat and Work} \label{sec:SIHW}
As established by statistical physics \cite{balian2007}, 
{\it work} is associated with changes in the system energy levels
$\hat H \rightarrow \hat H + \Delta \hat H$, 
and {\it heat} to variation of populations 
$\hat \rho \rightarrow \hat \rho + \Delta \hat \rho$.  
This dichotomy is naturally accomplished by the \hy{ME} (\ref{eq:mastereq}) 
in terms of the power $\mathcal P$ and of the current $\mathcal J$, 
both in (\ref{eq:powcurr}), 
since $d(\hat \rho \hat H) = \hat \rho \, d\hat H + d\hat \rho \, \hat H$. 
However, 
despite the covariance of the First Law in (\ref{eq:TransfmeanH}),  
the dichotomy is not so clear when we consider gauge transformations, 
and the reason is the acquired contributions 
for the power $\mathcal P'$ in (\ref{eq:gPower}) 
and for the current $\mathcal J'$ in (\ref{eq:gCur}).
In what follows, 
we will explore their physical meaning and 
their relations with this statistical interpretation.

The component ${\mathcal C}_{\! \delta \! \hat H}$ 
in (\ref{eq:gCur}) is a 
remarkable consequence of gauge transformations.  
The origin of this piece, defined in (\ref{eq:curpowdeltaH}), 
is fundamentally related to the change in the energy-eigenstates%
\footnote{From the spectral decomposition 
$\hat H = \sum_n E_n \ket{n}\!\bra{n}$, 
the variation of the energy has 
a contribution from the eigenvalues 
and another from the eigenvectors: 
$d \hat H = 
\sum_n d{E}_n \ket{n}\!\bra{n} + 
       E_n d[\ket{n}\!\bra{n}]$.} 
by the gauge transformation, 
it is thus a genuine quantum effect without a classical counterpart. 
Not for nothing, it is a power contribution
that appears in the current (\ref{eq:gCur}) 
or a power piece provided by the interaction with the environment. 
In detail, 
if we consider a transformation in (\ref{eq:GaugeTrans}), 
which preserves the energy eigenstates, 
{\it i.e.}, one gauge such that $[\hat H',\hat H ]=0$, 
then ${\mathcal C}_{\! \delta \! \hat H} = 0$. 
Otherwise, if $[\hat H',\hat H ] \ne0$, 
the gauges change the eigenstates, 
which is the origin of ${\mathcal C}_{\! \delta \! \hat H}$.  
The discrimination of a term directly 
in the First Law (\ref{eq:firstlaw}) 
related to eigenvectors variation 
is performed in works \cite{bernardo2020,ahmadi2022}. 

If the \hy{ME} has operators satisfying 
$[ \hat{L}_\mu , \hat H ] = 0, \forall \mu$, 
all the individual currents \eqref{eq:LindCurr} are null, 
consequently $\mathcal J = 0$ in (\ref{eq:sumcurr}).  
The Lindblad operators, 
which are diagonal in the energy eigenbasis, 
do not promote energy transitions, 
{\it i.e.} 
$\hat L_\mu \ket{n}$ belongs to the same ray as 
the energy eigenket $\ket{n}$, 
which is the reason behind $\text{Tr}[\mathfrak{D}(\hat \rho) \hat H] =0$, 
see (\ref{eq:powcurr}).
This can also be seen in writing 
$\text{Tr}[\mathfrak{D}(\hat \rho) \hat H] = \langle \mathfrak{D}^+(\hat H) \rangle$, 
for the adjoint operator $\mathfrak{D}^+$ \cite{BreuerPetr2002}. 
The above commutation relations are equivalent to $\mathfrak{D}^+(\hat H) = 0$, 
which annihilates the heat contribution $d\hat \rho \, \hat H$ 
to the mean energy variation. 
Besides, the only gauge contribution to $\mathcal J'$ in (\ref{eq:gCur}) 
is the current ${\mathcal J}_{\! \delta \! \hat H}$, 
since ${\mathcal C}_{\! \delta \! \hat H} = 0$ 
due to $[ \hat H', \hat H] = [ \delta\! \hat H, \hat H] = 0$ 
for $\delta\! \hat H$ in (\ref{eq:deltaH}). 

In a system described by a \hy{ME} with 
Lindblad operators such that 
$[ \hat{L}_\mu , \hat{L}_\nu ] = [ \hat{L}_\mu , \hat{L}_\nu^\dag ] = 0, 
\forall \mu, \nu$,
from Eq.\eqref{eq:deltaH}, 
$[\hat L_\mu, \delta \! \hat H] = 
[\hat L_\mu^\dag, \delta \! \hat H] = 0$, $\forall \mu$,  
then ${\mathcal J}_{\! \delta \! \hat H} = 0$ in (\ref{eq:curpowdeltaH}) 
by the same physical reason of the nullity of $\mathcal J$.  
In this case, the gauge contributions are the power in (\ref{eq:gPower}) 
and ${\mathcal C}_{\! \delta \! \hat H}$ in (\ref{eq:gCur}). 
Even when the system in question further satisfies 
$[\hat L_\mu, \hat H] = 0$, $\forall \mu$,
the power $\mathcal P'$ is still given by (\ref{eq:gPower}) and 
the system is still gauge-dependent, despite 
${\mathcal J}_{\! \delta \! \hat H} = 
 {\mathcal C}_{\! \delta \! \hat H} = 0$. 
In Sec.\ref{sec:ExPDM}, as an example,
we present a system satisfying 
all these commutation relations. 
Consider an {\it autonomous system}, 
the one in which the evolution is governed by (\ref{eq:mastereq}) 
with time-independent operators:  
$\partial_t \hat H = \partial_t \hat L_\mu = 0, \forall \mu$.  
For this class of systems, which contains the Lindblad \hy{ME},
the power is always null, $\mathcal P = 0$, see Eq.(\ref{eq:powcurr}).
However, $\mathcal P'$ in (\ref{eq:gPower}) may not be, 
due to the temporal dependence of the parameters in (\ref{eq:deltaH}).
Even for a gauge described by time-independent functions 
$\{U_{\mu\nu},\gamma_\mu,\phi\}$, 
where now it is also true that $\mathcal P' = 0$ due to 
$\partial_t \delta\! \hat H = 0$,
the current (\ref{eq:gCur}) is not invariant, 
as well as the First Law, since 
${\mathcal J}_{\! \delta \! \hat H}$ and 
${\mathcal C}_{\! \delta \! \hat H}$
in (\ref{eq:curpowdeltaH}) does not vanish 
for an autonomous system. 

Even the thermodynamics of a non-interacting system is 
gauge-dependent. 
If $\hat L_\mu = 0, \, \forall \mu$, the \hy{ME} (\ref{eq:mastereq}) 
becomes the Liouville-von Neumann unitary evolution
\begin{equation}\label{eq:LvNeq}
\partial_t \hat \rho = 
{\mathfrak U}(\hat \rho) = 
\tfrac{i}{\hbar}  [ \hat \rho , \hat H ], 
\end{equation}
which warrants $\mathcal J = 0$, 
as expected for an isolated system, see Eq.(\ref{eq:powcurr}). 
The addition of a dynamical phase, 
$\hat H \rightarrow \hat H' = \hat H + \phi(t)$,
is a gauge-symmetry of the unitary evolution, 
which generates a power contribution for the system energy:
\[
\partial_t \langle \hat H' \rangle = \mathcal P'
= \mathcal P + \partial_t \phi,            
\]
according to Eq.(\ref{eq:gFirstLaw2}) with 
$\delta\!\hat H = \phi(t)$, see Eq.(\ref{eq:deltaH}), 
and 
${\mathcal J}_{\! \delta \! \hat H} = 
 {\mathcal C}_{\! \delta \! \hat H} =0$, 
due to $\hat L_\mu = 0$, see Eq.(\ref{eq:curpowdeltaH}).
Note that this very same gauge symmetry is also present in 
the dynamical map (\ref{eq:dynmap}) through the unitary evolution 
associated with the global Hamiltonian $\hat H_{\rm SE}$. 
%

\section{Thermodynamical Invariance}\label{sec:ThermInv}
In the scenario described so far, 
gauges influence the thermodynamics of a system, 
raising the question about the existence of a possible invariant 
thermodynamical behavior and under what conditions 
this invariance is manifested.

Following the description in (\ref{eq:gCur}), 
the invariance of the Lindblad total current is attained when 
\begin{equation}\label{eq:Invcurr}
{\mathcal J}' = {\mathcal J} \Longleftrightarrow
{\rm Tr}(\partial_t \hat \rho \, \delta\! \hat H ) 
= {\mathcal J}_{\! \delta \! \hat H} + 
{\mathcal C}_{\! \delta \! \hat H} = 0,     
\end{equation}
and from (\ref{eq:gPower}) the invariance 
of the Lindblad power is such that
\begin{equation}\label{eq:Invpow}
{\mathcal P}' = {\mathcal P} \Longleftrightarrow
\langle\partial_t \delta\! \hat H \rangle = 0.    
\end{equation}
In this way, the First Law in (\ref{eq:firstlaw}) 
is invariant provided the conditions 
of invariance of the current in (\ref{eq:Invcurr}) 
and of power in (\ref{eq:Invpow}) are met, as it should be.  
However, 
the condition for the invariance of the mean energy (or its rate) is   
\begin{equation}\label{eq:Invmeanen}
\langle\hat H'\rangle = 
\langle\hat H \rangle \Longleftrightarrow
\langle\delta\!\hat H\rangle = 0 \Longrightarrow 
\partial_t \langle\delta\!\hat H\rangle = 0,    
\end{equation}
which is only a necessary condition for both current and power invariances, 
since $\langle\delta\!\hat H\rangle = 0$ does not mean (\ref{eq:Invcurr}) 
neither (\ref{eq:Invpow}).

Let us give a closer look to condition (\ref{eq:Invpow}). 
Using the definition in (\ref{eq:deltaH}), one writes
\begin{equation*}
\begin{aligned}
\langle\partial_t \delta\! \hat H \rangle &= 
 {\rm Im}
\langle g_{\mu\nu} \dot\gamma_\mu^\ast U_{\nu\kappa} {\hat L}_\kappa \rangle + 
 {\rm Im}
\langle g_{\mu\nu} \gamma_\mu^\ast\dot U_{\nu\kappa} {\hat L}_\kappa \rangle \\ 
&+ {\rm Im}
\langle g_{\mu\nu} \gamma_\mu^\ast U_{\nu\kappa} \partial_t{\hat L}_\kappa \rangle  
+ \partial_t\phi; 
\end{aligned}
\end{equation*}
%
consequently, 
the invariance in (\ref{eq:Invpow}) for any gauge (\ref{eq:GaugeTrans}), 
\emph{i.e.}, for arbitrary functions 
$\gamma_\mu$, $U_{\!\mu\nu}$, and $\phi$, becomes
\begin{equation}\label{eq:invcond1}
\langle\partial_t \delta\! \hat H \rangle = 0 \Longleftrightarrow
\partial_t\phi = \langle {\hat L}_\mu \rangle = 
\langle \partial_t{\hat L}_\mu \rangle 
= 0, \,\,\, \forall \mu. 
\end{equation}

The invariance condition in (\ref{eq:Invcurr}), 
using $\delta \! \hat H$ from (\ref{eq:deltaH}), becomes 
$
{\rm Im}[ 
g_{\mu\nu} \gamma_\mu^\ast U_{\nu\kappa}
          {\rm Tr}( {\hat L}_\kappa \partial_t \hat \rho) ] = 0
$. 
For arbitrary gauge functions, 
this is the same as 
\begin{equation}\label{eq:invcondaux}
{\rm Tr}( {\hat L}_\mu \partial_t \hat \rho) = 
\partial_t \langle \hat L_\mu\rangle - 
\langle \partial_t \hat L_\mu\rangle  = 0, \,\,\, \forall \mu;  
\end{equation}
finally, one can write
\begin{equation}\label{eq:invcond2}
 {\mathcal J}_{\! \delta \! \hat H} +
{\mathcal C}_{\! \delta \! \hat H} = 0 \Longleftrightarrow 
\partial_t \langle \hat L_\mu\rangle = 
\langle \partial_t \hat L_\mu\rangle, \,\,\, \forall \mu. 
\end{equation}

The statements in (\ref{eq:invcond1}) and in (\ref{eq:invcond2}) are, 
in principle, independent of each other.  
However, to obtain a consistent description for the First Law, 
expressed as (\ref{eq:firstlaw}), 
where each term (energy, power, and current) 
is invariant under all possible gauges of the \hy{ME}, 
it is necessary and sufficient to have
\begin{equation} \label{eq:invCond}
\partial_t\phi = \langle {\hat L}_\mu \rangle = 
\langle \partial_t{\hat L}_\mu \rangle 
= 0, \,\,\, \forall \mu.
\end{equation}
For an autonomous system, these reduce to
$\partial_t \phi = \langle {\hat L}_\mu \rangle = 0$.

The standard First Law (\ref{eq:firstlaw}) 
works well for invariant systems since  
$\mathcal J$ and $\mathcal P$ are not affected by 
any possible gauge of the system.  
Notwithstanding, 
the invariance conditions (\ref{eq:invCond}) are very restrictive: 
they must be satisfied at any time, 
including system initial state.    
We are thus led to conclude that      
the thermodynamics of the vast majority of systems is influenced 
by the gauge symmetries of the \hy{ME} 
and requires a consistent interpretation 
based on Eqs.(\ref{eq:gCur}), 
(\ref{eq:gPower}), and (\ref{eq:TransfmeanH}).
We will provide this interpretation in Sec.\ref{sec:GaugeTherm},   
but before, we will spend some words discussing 
those conditions and presenting an important 
class of invariant systems. 

\subsection{Invariance Conditions}
Here we will explore the physical meaning of the invariance 
conditions in Eq.(\ref{eq:invCond}) and its relation with the system thermodynamics. 

A time-dependent phase $\phi(t)$ would continuously change 
the energy of the system. 
Since it appears as part of the Hamiltonian through $\delta\!\hat H$, 
see Eq.(\ref{eq:deltaH}), it will be properly quantified by work. 
Therefore, $\partial_t \phi = 0$ in (\ref{eq:invCond}) 
is related to the invariance of power, see (\ref{eq:Invpow}). 
For a constant phase, 
all the system energy levels will be shifted as a mere 
redefinition of the ground state energy, as in the Schrödinger equation.
We will return to this point in Sec.\ref{sec:UnD-NIntSys}.

Condition $\langle {\hat L}_\mu \rangle=0$ implies 
that the energy contribution $\delta\!\hat H$ promoted by any gauge, 
see Eqs.(\ref{eq:GaugeTrans}) and (\ref{eq:deltaH}),  
does not impact the internal energy of the system on average, 
$\langle \hat H' \rangle = \langle \hat H\rangle$.  
Assuming that 
$\hat \rho$ is expanded in the eigenbasis 
of the system Hamiltonian $\hat H$, 
and that ${\hat L}_\mu$ stands for a projector operator 
at some definite level of this Hamiltonian, 
then the mean-value $\langle {\hat L}_\mu \rangle$ 
will be the time-dependent occupation probability of this level --- 
Condition $\langle {\hat L}_\mu \rangle=0$ 
will be fulfilled for a vanishing occupation probability, 
which can occur solely for specific initial states. 
Furthermore, if ${\hat L}_\mu$ stands for some 
energy transition, \emph{i.e.}, it is a jump operator, 
then $\langle {\hat L}_\mu \rangle$ becomes a probability amplitude 
--- The invariance condition requires the vanishing of this amplitude 
for all time, which again depends on the particular initial state.

For a generic time-dependent Lindblad operator, 
the invariance conditions must include 
$\langle \partial_t{\hat L}_\mu \rangle = 0$.
From Eq.(\ref{eq:gPower}) and $\delta\!\hat H$ in (\ref{eq:deltaH}), 
this condition ensures the power invariance due to 
$\langle \partial_t{\delta\!\hat H} \rangle = 0$. 
Note that, 
if the energy and power are invariant, 
logically the current will either be, %
mathematically this is expressed in 
Eqs.(\ref{eq:invcondaux}) and (\ref{eq:invcond2}) 
when the former conditions, $\partial_t \phi = \langle {\hat L}_\mu \rangle=0$, 
are valid.

Among many possibilities for the temporal dependence of the Lindblad operators, 
the particular case $\hat L_\mu = \omega_\mu(t) \hat A_\mu$, 
for a time-independent $\hat A_\mu$, 
turns the third invariance condition into 
$\langle \partial_t{\hat L}_\mu \rangle = \dot{\omega}_\mu \hat A_\mu = 0$,  
which is analogous to the previous discussion in terms of probabilities. 
Actually, even for a generic time-dependence, 
when condition 
$\langle {\hat L}_\mu \rangle=0$ 
is met, the condition 
$\langle \partial_t{\hat L}_\mu \rangle = 0$ 
is subsidiary 
since applying (\ref{eq:invcondaux}) implies 
$\langle \partial_t \hat \rho {\hat L}_\mu \rangle = 0$,
which, for an infinitesimal evolution, shows that   
$\langle \hat\rho(t+dt) {\hat L}_\mu \rangle = 
\langle \hat\rho(t) {\hat L}_\mu \rangle$, 
{\it i.e.}, $\langle \hat L_\mu \rangle$ is constant.
Although subsidiary, this condition is necessary 
for the invariance of the current through (\ref{eq:invcondaux}). 

\subsection{Quantum Detailed Balance}
Although specific, 
a significant property of some \hy{ME}s is their ability 
to drive the system toward thermal equilibrium. 
In analogy with classical stochastic processes, 
this property is mathematically expressed by the so-called 
Quantum Detailed Balance Conditions 
\hypertarget{QDBC}{(QDBC)}\cite{QDBCfin}. 

For a time-independent Hamiltonian $\hat H_\text{S}$ 
with non-null Bohr frequencies $\omega_1,...,\omega_n$ 
and an autonomous Lindblad \hy{ME}, 
governing the evolution of the system,  
composed by Lindblad operators $\hat A_\mu$ and 
by another Hamiltonian $\hat H$, the \hy{QDBC} are 
\begin{equation*}
\begin{aligned}
& \hypertarget{(a)}{\text{(a)}}   \; [ \hat H_\text{S} , \hat H ] = 0; \\    
& \hypertarget{(b)}{\text{(b)}} \; 
\hat U_\text{S}^\dag \hat A_{\mu} \hat U_\text{S} = 
{\rm e}^{-i \omega_\mu t } \hat A_{\mu}, \,\,\, 
\hat U_\text{S}:= {\rm e}^{-\frac{i}{\hbar} \hat H_\text{S} t};  \\
& \hypertarget{(c)}{\text{(c)}} \; \hat A_{\mu + n} =
{\rm e}^{ -\tfrac{1}{2}\beta \hbar \omega_\mu} \hat A_{\mu}^\dag,  \,\,\, 
1 \le \mu \le n.     
\end{aligned}
\end{equation*}
The physical consequence of these conditions will be described in a while, 
just after we show the main result of this part: 
a \hy{ME} satisfying these \hy{QDBC} 
is thermodynamically invariant under gauges transformations. 

Under condition \hy{(c)}, 
the generator $\mathfrak D$ in (\ref{eq:genME}) becomes 
%
%
\[
\tilde{{\mathfrak D}}(\hat \rho) = 
- \tfrac{1}{2\hbar} \sum_{\mu = 1}^{n}  
( \{ \hat{A}_{\mu}^\dag \hat{A}_{\mu} , \hat \rho  \}
- 2 \hat{A}_{\mu} \hat \rho \hat{A}_{\mu}^\dag )
(1-{\rm e}^{-\hbar\beta \omega_\mu});  \\
\]
note that condition \hy{(c)} imposes $M = 2n$ 
in (\ref{eq:mastereq}). 

Using condition \hy{(b)} in the above generator $\tilde{\mathfrak D}$, 
it is possible to show that 
$\hat U_\text{S}\tilde{{\mathfrak D}}(\hat \rho)\hat U_\text{S}^\dag = 
\tilde{{\mathfrak D}}(\hat U_\text{S} \hat \rho \hat U_\text{S}^\dag)$;  
from condition \hy{(a)}, immediately one has
$\hat U_\text{S}{\mathfrak U}(\hat \rho)\hat U_\text{S}^\dag = 
{\mathfrak U}(\hat U_\text{S} \hat \rho \hat U_\text{S}^\dag)$  
for the generator $\mathfrak{U}$ in (\ref{eq:genME}).
Thus, the \hy{ME} 
$\partial_{t} \hat \rho = {\mathfrak U}(\hat \rho) + 
\tilde{{\mathfrak D}}(\hat \rho)$ 
satisfies
$\hat U_\text{S} (\partial_t \hat \rho) \hat U_\text{S}^\dag = 
\partial_{t} (\hat U_\text{S} \hat \rho \hat U_\text{S}^\dag)$, 
which implies $[\hat \rho,\hat H_\text{S}]=0$ and 
$\hat U_\text{S} \hat \rho \hat U_\text{S}^\dag = \hat \rho$. 

Since $[\hat \rho,\hat H_\text{S}]=0$, 
$\partial_t \hat A_\mu = 0$, 
and $\partial_t \omega_\mu = 0$, 
condition \hy{(b)} is equivalently written as \cite{toscano2022,QDBCfin} 
\[
\hat\chi_\text{S}^{+} \hat A_{\mu} \hat\chi_\text{S}^{-} = 
{\rm e}^{-\beta \hbar \omega_\mu } \hat A_{\mu} , 
\,\,\,
\hat\chi_\text{S}^{\pm} := \exp[\pm \beta \hat H_\text{S}], 
\]
which is nothing but a Wick-rotated version of \hy{(b)}. 
With this in hands, 
the cyclicity of the trace readily gives 
\[
\langle \hat A_\mu \rangle =  {\rm Tr}(\hat \rho \hat A_\mu) =
{\rm Tr}(\hat\chi_\text{S}^{-} \hat \rho \hat\chi_\text{S}^{+} \hat A_\mu) = 
{\rm e}^{-\beta\hbar\omega_\mu}\langle \hat A_\mu \rangle,    
\]
from where 
$({\rm e}^{-\beta\hbar\omega_\mu} - 1)\langle \hat A_\mu \rangle = 0$. 
As required by the \hy{QDBC} $\omega_\mu \ne 0$, 
thus $\langle \hat A_\mu \rangle = 0, \forall \beta \ne 0$. 

Instead of using $\hat\chi_\text{S}^{\pm}$, 
the same steps for the (unrotated) unitary operator $\hat U_\text{S}$ 
would give $({\rm e}^{-i\omega_\mu t} - 1)\langle \hat A_\mu \rangle = 0$,
which certifies that the instant $t = 0$ 
corresponds to the value $\beta = 0$. 
Finally, we can conclude that 
$\langle \hat A_\mu \rangle = 0, \forall t \ne 0$, 
which is the requisite in (\ref{eq:invCond}) 
for the thermodynamical invariance of an autonomous system, 
except for two facts: 
the initial state $\hat \rho(t=0)$ 
and the condition over $\phi$ ---  
the discussion about the phase and its meaning in (\ref{eq:invCond})
will be postponed to Sec.\ref{sec:UnD-NIntSys}, 
till then, it will be taken for granted.

The invariance of currents, power, and energy of 
a system satisfying \hy{QDBC} for the whole evolution will only 
be valid if its initial state is such that
$\langle \hat A_\mu \rangle = 0$. 
If not, the thermodynamical invariance
can possibly occur only for fixed points of the \hy{ME}, 
these can be an asymptotic state, or even thermal equilibrium states,   
which move us back to the physical consequence of the \hy{QDBC}. 

For some dynamical semigroups,
the \hy{QDBC} are necessary and sufficient conditions to 
\begin{equation*}
\lim_{t \to \infty} \hat \rho(t) = 
\hat\chi_\text{S}^{-}/{\rm Tr}(\hat\chi_\text{S}^{-}) = 
\text{e}^{-\beta \hat H_\text{S}} / {\rm Tr}(\text{e}^{-\beta \hat H_\text{S}}),  
\end{equation*}
which means that the asymptotic state of the \hy{ME}, 
also a fixed point,
is the Gibbs state of the Hamiltonian $\hat H_\text{S}$. 
As far as we know, 
this is only proved for dynamical semigroups 
of finite-dimensional quantum systems \cite{QDBCfin} 
and for Gaussian dynamical semigroups of 
continuous variables systems \cite{toscano2022}
--- a very small set in the universe of \hy{ME}s. 
Nonetheless, a generic \hy{ME} 
that does not belong to this set can satisfy the \hy{QDBC}
and will be thermodynamically invariant throughout 
the whole evolution for a suitable choice of 
the initial state, as proved. 

Examples of Markovian systems satisfying the \hy{QDBC} 
are the infamous quantum optics master equations for bosonic systems
and discrete systems, including the decay of a two-level atom, 
for instance, see Section 3.4 of \cite{BreuerPetr2002}.

\section{Gauge-Dependent Thermodynamics}\label{sec:GaugeTherm}
From the dynamical point of view, 
it is impossible to discriminate what would be the proper gauge of the evolution 
since the dynamics is governed by a gauge-invariant \hy{ME}. 
The only plausible way to distinguish between gauges, 
or to determine the amount of work and heat in a certain process, 
is a choice of measurements to be performed on the system. 
The internal energy $\langle \hat H' \rangle$ can be determined after 
the measurement of the corresponding operator $\hat H'$ in Eq.(\ref{eq:GaugeTrans}),
associated with an infinite set of possible operators $\hat L'_\mu$,  
through the choice of $\delta\! \hat H$ in (\ref{eq:deltaH}), 
one piece of $\hat H'$.
A choice for the Hamiltonian is precisely the 
first step in modern tomographic methods \cite{TomogProc}.

Instead of considering the symmetries of the \hy{ME} as an impossibility to
determine the thermodynamical quantities correctly, 
we place our analysis on the classical interpretation ground: 
Each possible gauge defined by a choice of the functions 
$\{{\gamma_\mu,U_{\!\mu\nu}},\phi\}$ in (\ref{eq:GaugeTrans})  
represents a particular thermodynamical process (or path) with 
its own amount of heat and work. 

Contrasting with classical thermodynamics, 
the energy is not a ``state function'', 
it varies according to the gauge (a thermodynamical path), 
$\langle\hat H'\rangle = \langle\hat H \rangle + \langle\delta\!\hat H\rangle$. 
In our scenario, state functions are gauge-invariant quantities, 
which depend solely on the system state $\hat \rho$; 
therefore, any function with domain in the space of density matrices 
is invariant, \emph{e.g.}, the von Neumann-entropy, 
measures of non-Markovianity based on divisibility of quantum maps, 
or on distinguishability of quantum states \cite{breuer2016}.

\begin{figure}[!tb]
\centering
\includegraphics[width=1.0\columnwidth]{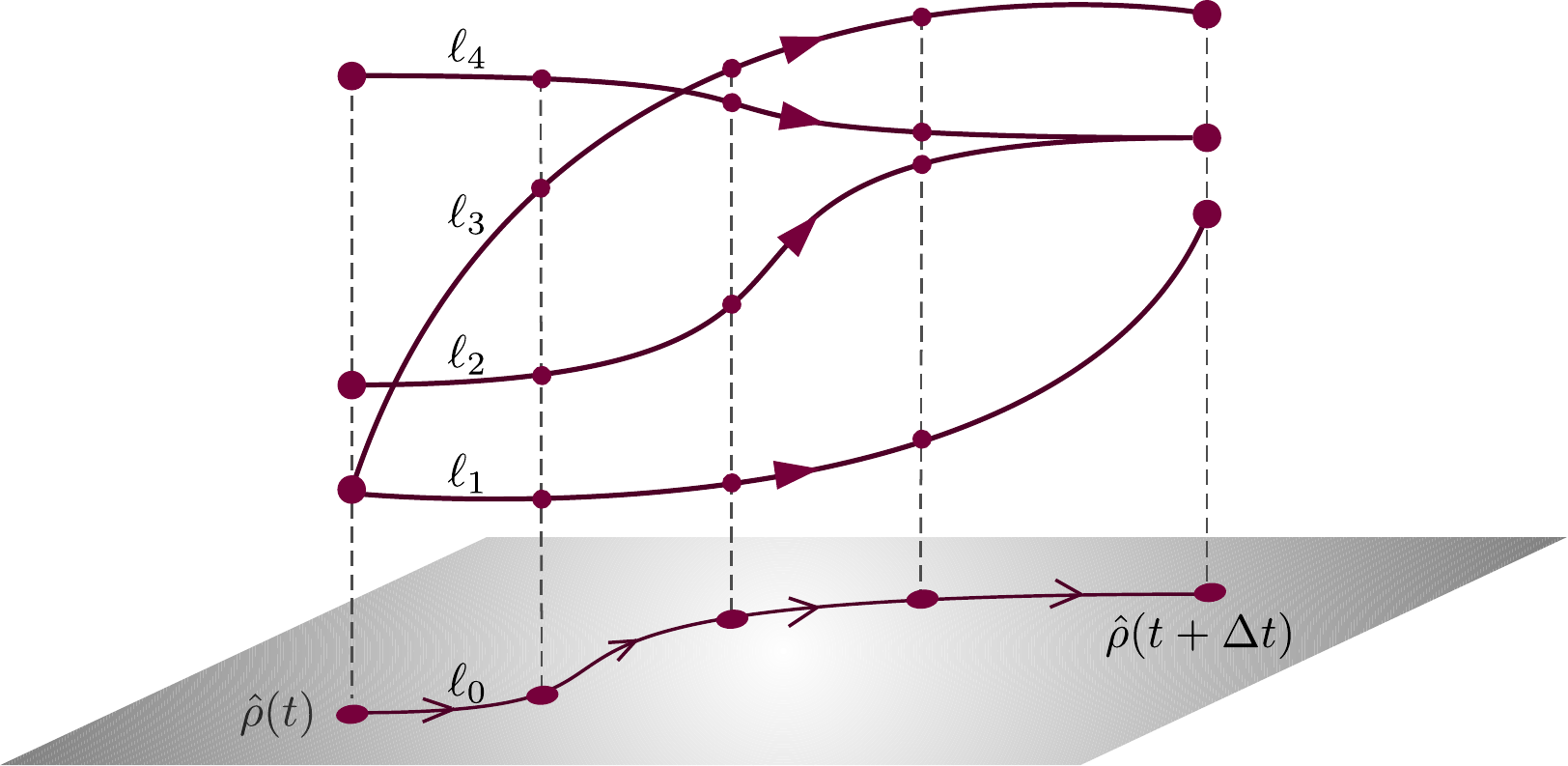} 
\caption{%
Schematic Representation of Thermodynamical Paths.
Each of curves 
$\ell_1$, $\ell_2$, $\ell_3$, and $\ell_4$
represents a specific thermodynamical process, 
with its amount of heat and work. 
The covariance of the mean energy rate in (\ref{eq:TransfmeanH}) 
ensures the validity of the First Law along each path 
since the sum of heat and work
gives the variation of the (gauge-dependent) mean energy. 
These paths differ by the gauge parameters 
$\{\gamma_\mu,U_{\!\mu\nu},\phi\}$  
and are projected into a unique curve $\ell_0$ in the state space, 
representing the system evolution and its invariance under 
gauge transformations. 
A gauge can even change during the evolution
due to the temporal dependence of 
$\{\gamma_\mu,U_{\!\mu\nu},\phi\}$,
a situation depicted by the crossing curves. }       \label{fig:Gauge}                                                                     
\end{figure}

The thermodynamics embodied in the system evolution 
is only properly described by \hy{ME} written as 
$\partial_t \hat \rho = 
\mathfrak{L}'(\hat \rho)= 
\mathfrak{U}'(\hat \rho) + \mathfrak{D}'(\hat \rho)$,
for the generators in (\ref{eq:gGenME}), 
where $\hat H'$ and $\hat{L}_\mu'$ are the operators in (\ref{eq:GaugeTrans}), 
which are defined by the set
\begin{equation}\label{eq:pathl}
\ell = 
\{\hat H,\hat L_\mu,\gamma_\mu,U_{\!\mu\nu},\phi ; \mu,\nu = 1,...,M\}, 
\end{equation}
the thermodynamical path or process. 
The projection of all paths sharing 
the same operators $\hat H$ and $\hat L_\mu$ 
is the path 
\begin{equation}\label{eq:pathl0}
\ell_0 = \{\hat H,\hat L_\mu; \mu = 1,...,M\},
\end{equation}
corresponding to the system evolution 
through the invariant \hy{ME}, 
as pictorially represented in Fig.\ref{fig:Gauge}. 
Formally, the set 
$\ell_0$ in (\ref{eq:pathl0})
is an equivalence class of invariant dynamics under 
the group of transformations in (\ref{eq:GaugeTrans}).  
This class contains all sets $\ell$ defined by Eq.(\ref{eq:pathl}).

The heat $Q_\ell$ and the work $W_\ell$ 
along a path $\ell$, thus gauge-dependent, 
are determined by ${\mathcal J}'$ in (\ref{eq:gCur}) and 
${\mathcal P}'$ in (\ref{eq:gPower}): 
%
%
\begin{equation}\label{eq:heatworkc}
\begin{aligned}
Q_\ell  &= \int_\ell {\mathcal J}' dt =
\int_\ell 
( {\mathcal J} + {\mathcal J}_{\! \delta \! \hat H} + 
                 {\mathcal C}_{\! \delta \! \hat H}) dt,   \\
W_\ell &= \int_\ell {\mathcal P}' dt =
\int_\ell 
( {\mathcal P}  + 
  \langle \partial_t \delta\! \hat H \rangle          ) dt, 
\end{aligned}
\end{equation}
for 
$\mathcal P$ 
and 
$\mathcal J$ 
both defined in (\ref{eq:powcurr}), 
and  
${\mathcal J}_{\! \delta \! \hat H}$ 
and 
${\mathcal C}_{\! \delta \! \hat H}$ 
both defined in (\ref{eq:curpowdeltaH}).
Needless to say, 
$Q_\ell + W_\ell = \Delta \langle \hat H' \rangle$ 
is the integral representation of the First Law in (\ref{eq:TransfmeanH}). 

For thermodynamical-invariant systems,  
the ones satisfying the invariance conditions (\ref{eq:invCond}), 
all thermodynamical paths (\ref{eq:pathl}) collapse into 
${\ell_0}$, the path identified with the system evolution, 
thus $Q_{\ell} = Q_{\ell_{0}}$ and $W_{\ell} = Q_{\ell_{0}}$, $\forall \ell$, 
with
\begin{equation}\label{eq:heatworkc0}
Q_{\ell_{0}} = \int_{\ell_{0}}\!\! {\mathcal J}dt, \,\,\, 
W_{\ell_{0}} = \int_{\ell_{0}}\!\! {\mathcal P}dt, \,\,\, 
\Delta \langle \hat H \rangle = Q_{\ell_{0}} + W_{\ell_{0}}.
\end{equation}
For non-invariant systems, the differences 
\begin{equation}\label{eq:heatworkVar}
\begin{aligned}
Q_{\ell} - Q_{\ell_{0}} &= 
\int_{\ell_{0}} ( {\mathcal J}_{\! \delta \! \hat H} + 
{\mathcal C}_{\! \delta \! \hat H} ) dt, \\
W_{\ell} - W_{\ell_{0}} & =  
\int_{\ell_{0}} \langle \partial_t \delta\! \hat H \rangle dt, 
\end{aligned}
\end{equation}
are the gauge contributions for the heat and work and, 
of course, are related through 
$\Delta \langle \hat H' \rangle - \Delta \langle \hat H \rangle = 
\Delta\langle\delta \! \hat H\rangle$. 
%

%
%

Path $\ell_0$ is not special compared to the other paths $\ell$, 
it is actually only a didactic reference, 
a choice for operators of the system, 
which can be anyone in (\ref{eq:GaugeTrans}) 
or any path (\ref{eq:pathl}). 
The only fixed quantity is the generator $\mathcal L$ of the \hy{ME}
since it is the same for any path.
Once $\ell_0$ is chosen, 
all the other paths, and all thermodynamical quantities, 
will be related to this one by 
gauge transformations (\ref{eq:GaugeTrans}). 

In the end, we must say that there is nothing new regarding 
a path-dependence in quantum thermodynamics   
since two quantum states can be connected 
by the temporal evolution ruled by different \hy{ME}s; 
in our description, these \hy{ME}s are associated with 
different curves $\ell_0$ in the state space (equivalence classes), 
see Fig.\ref{fig:Gauge}.
Our interpretation resides on the association of different 
paths $\ell$ to the same state evolution.

\subsection{Unitary Dynamics and Non-interacting Systems}\label{sec:UnD-NIntSys}
The analysis of this simple situation may be enlightening and paves the way 
to discussing the remaining point about the invariance condition 
$\partial_t \phi =0$ in (\ref{eq:invcond1}). 

Regarding the unitary dynamics ruled by (\ref{eq:LvNeq}), 
mathematically the phase addition 
$\hat H \rightarrow \hat H' = \hat H + \phi(t)$ 
is a ${\rm U}(1)$-symmetry 
of the Liouville-von Neumann equation and a thermodynamical path 
$\ell = \{\hat H, \phi\}$ 
is the one-dimensional manifold of the unitary operator 
$\hat U'_t$, 
the solution of the Schrödinger equation 
$i\hbar \partial_t\hat U'_t = \hat H'\hat U'_t$. 
Two paths, say ${\ell }_1 = \{\hat H, \phi_1\}$ and 
$\ell_2 = \{\hat H, \phi_2\}$, 
are members of the equivalence class 
$\ell_0 = \{\hat H\}$, 
the manifold of the unitary operator satisfying 
$i\hbar \partial_t\hat U_t = \hat H\hat U_t$. 

Of course, the phase $\phi(t)$ is a global and instantaneous 
shift of the energy levels, 
which does not change the system dynamics (\ref{eq:LvNeq}), 
however, it does cause a change in system energy, 
thermodynamically accounted by the work in (\ref{eq:heatworkc})
with $\delta\!\hat H = \phi(t)$, ${\mathcal C}_{\! \delta \! \hat H} =0$, 
and $\mathcal P = \langle \partial_t \hat H\rangle$. 
As a trivial example, 
the phase $\phi(t) = \tfrac{1}{2}\hbar \omega(t)$
is a power source appearing in the 
time-dependent Hamiltonian
$\hat H = \hbar\omega(t) \hat n + \phi(t)$ of the harmonic oscillator; 
some thermodynamical 
properties of this system were described in \cite{HOAppl}.

We now return to the point regarding the phase and
the thermodynamical invariance.  
Whereas $\phi$ has its origin in a gauge transformation, 
the condition $\partial_t \phi = 0$ in Eq.(\ref{eq:invcond1}), 
and also in Eq.(\ref{eq:invCond}), is awkward.   
Nevertheless, $\phi$ appears as a source of power for a system 
governed by a \hy{ME} through $\delta\!\hat H$, 
see Eqs.(\ref{eq:gPower}) and (\ref{eq:deltaH}),  
in the same way as it appears in the unitary evolution, 
where the power is unequivocally defined 
and addition of phases simply describes the gauge group. 
Strictly speaking, there is no thermodynamical invariance at all 
if one is free to add a time-dependent phase to the Hamiltonian.  
As in the unitary dynamics, 
the phase will change the energy of a system governed by a 
\hy{ME} (\ref{eq:mastereq}). 
Besides $\langle {\hat L}_\mu \rangle = 
\langle \partial_t{\hat L}_\mu \rangle 
= 0$, imposing the invariance condition for $\phi$ in (\ref{eq:invCond}) 
is actually a restriction of the allowed gauge functions.
%
\subsection{Analysis of Particular Processes} \label{sec:AnParProc}
Among all possible gauges, 
it is interesting to pinpoint some particular thermodynamical features associated 
with specific processes for generic non-invariant systems.   
\vspace{0.2cm}

\noindent $(i)$ \emph{Energy Preservation}: 
The internal energy of a system is not gauge-invariant, 
$\langle\hat H'\rangle = \langle\hat H \rangle + \langle\delta\!\hat H\rangle$,  
see Eq.(\ref{eq:gFirstLaw2}).
Notwithstanding, there are many paths $\ell$ in (\ref{eq:pathl}) 
such that $\langle\hat H'\rangle = \langle\hat H \rangle$: 
these are the ones such that 
$\langle\delta\!\hat H\rangle = 0$, of course%
\footnote{
The system energy invariance by 
$\langle\delta\!\hat H\rangle = 0$ 
is the necessary condition expressed in (\ref{eq:Invmeanen}).}.  

Considering the \hy{ME} with the transformed generators (\ref{eq:gGenME}), 
the suitable choice 
$\gamma_\mu = U_{\!\mu\nu} \langle \hat L_\nu\rangle$ and $\phi =0$ 
designs a path $\ell$ such that
$\delta\!\hat H =  \tfrac{ g_{\mu\nu} }{2i} 
          [\langle \hat L_\mu^\dagger\rangle \hat L_\nu - 
           \langle \hat L_\mu\rangle  \hat L_\nu^\dagger]$, 
according to (\ref{eq:deltaH}). 
Consequently, $\langle\delta\!\hat H\rangle = 0$ and 
$\langle\hat H'\rangle = \langle\hat H \rangle$.
For the chosen gauge, 
the current is given by (\ref{eq:gCur}) and the power by (\ref{eq:gPower})
for the above $\delta\!\hat H$ and, following (\ref{eq:derivdeltaH}), 
these are related through 
${\mathcal J}_{\! \delta \! \hat H}  + {\mathcal C}_{\! \delta \! \hat H} 
= -\langle\partial_t \delta\! \hat H \rangle$.  
Note that this is true for any matrix $\bf U$, 
since the attained $\delta\!\hat H$ does not depend on it. 

For the same purpose, 
another path $\ell$ is determined by the choice  
\[
\phi =  - \tfrac{ g_{\mu\nu} }{2i} 
(\gamma_\mu^\ast U_{\nu\kappa}\langle \hat L_\kappa \rangle- 
 \gamma_\mu      U_{\nu\kappa}^\ast\langle\hat L_\kappa^\dagger\rangle),
\]
which, according to (\ref{eq:deltaH}), is such that 
$\langle\delta\!\hat H\rangle = 0$.  
Now Eq.(\ref{eq:gFirstLaw2}) gives
$\mathcal J + \mathcal P = \mathcal J' + \mathcal P'$ 
for any $\gamma_\mu$ and any $\bf U$. 

If in $\ell_0$, 
the work exactly balances heat, \emph{i.e.}, 
$\mathcal P = - {\mathcal J}$, 
the energy of the system following $\ell_0$ 
will be conserved, $\partial_t \langle \hat H \rangle = 0$, 
according to (\ref{eq:firstlaw}). 
Above designed gauges are specifically useful to preserve this conservation 
since another generic gauge transformation will generate a path with 
non-conserved internal energy $\langle\hat H'\rangle$ such that 
$\partial_t \langle\hat H'\rangle = \partial_t \langle\delta\!\hat H\rangle$, 
obtained from Eqs.(\ref{eq:derivdeltaH}) and Eqs.(\ref{eq:gFirstLaw2}) 
with $\mathcal P = - {\mathcal J}$. 

\vspace{0.2cm}

\noindent $(ii)$ \emph{Power Preservation:} 
Consider the following phase
\[
\phi = - \tfrac{ 1}{2i}{\int_0^t}  \langle \partial_\tau
(g_{\mu\nu}\gamma_\mu^\ast U_{\nu\kappa}\hat L_\kappa - 
 g_{\mu\nu}\gamma_\mu      U_{\nu\kappa}^\ast\hat L_\kappa^\dagger) \rangle d\tau,       
\]
which defines a path 
$\ell_1$ such that 
$\mathcal P' = \mathcal P$, see (\ref{eq:gPower}), 
since $\langle \partial_t \delta\! \hat H \rangle = 0$.  
The current will be the one in Eq.(\ref{eq:gCur}) with 
${\mathcal J}_{\! \delta \! \hat H} + {\mathcal C}_{\! \delta \! \hat H} = 
\partial_t \langle \delta\!\hat H \rangle$. 
Beyond that, one can choose a path $\ell_2$ 
for the same phase above and also take
$\gamma_\mu = U_{\!\mu\nu} \langle \hat L_\nu\rangle$ 
for a generic $\bf U$, 
then $\langle\delta\!\hat H \rangle = \phi$.  
However $\mathcal J' = \mathcal J + \partial_t\phi$ and 
$\langle\hat H'\rangle = \langle\hat H \rangle + \phi$ in $\ell_2$. 
Even in the particular case of an autonomous \hy{ME}, 
$\mathcal J' \ne \mathcal J$, despite $\mathcal P' = \mathcal P = 0$, 
since $\phi$ is in general a time-dependent function. 

\vspace{0.2cm}

\noindent $(iii)$ \emph{Current Preservation:} 
Since $\det {\bf U} \ne 0$, 
it is possible to choose $\gamma_\mu$ as a ``$\bf g$-orthogonal'' vector to 
$ {\rm Tr}[U_{\nu\kappa}\hat L_\kappa \partial_t\hat \rho ]$, such that 
$g_{\mu\nu} \gamma_\mu^\ast U_{\nu\kappa} 
{\rm Tr}[\hat L_\kappa \partial_t\hat \rho ] = 0$, 
according to which $\mathcal J' = \mathcal J$, 
as can be seen by inserting (\ref{eq:deltaH}) into (\ref{eq:Transfcur2}) 
and comparing with (\ref{eq:gCur}).
Additionally, 
by choosing a phase $\phi$ such that $\langle\delta\!\hat H\rangle=0$, 
like in $(i)$, one obtains $\langle\hat H'\rangle = \langle\hat H \rangle$
and thus $\mathcal P = \mathcal P'$. 
This choice for the functions $\{\gamma_\mu, \phi\}$ 
designs infinite paths $\ell \ne \ell_0$, see Eq.(\ref{eq:pathl}), 
in which all thermodynamical quantities are preserved, 
{\it i.e.}, they have the same value as in $\ell_0$. 

\vspace{0.2cm}

\noindent $(iv)$ \emph{Minimal Dissipation:} 
for systems described by 
finite-dimensional Hilbert spaces, 
the gauge fixed by $U_{\!\mu\nu} = \delta_{\mu\nu}$ and 
$\gamma_\mu = -{\rm Tr}(\hat L_\mu)$ 
ensures ${\rm Tr}(\hat L'_\mu) = 0$ in (\ref{eq:GaugeTrans}), 
which is the minimal dissipation condition postulated in \cite{colla2022}.  

\section{Entropy Production and Second Law} \label{sec:EntProd}
The von-Neumann entropy $\mathcal{S} := -\langle {\rm ln} \hat \rho \rangle$ 
is a state function, an exclusive function of the density operator, 
thus invariant under gauge transformations.  
According to (\ref{eq:mastereq}), its evolution is 
\begin{equation} \label{eq:entEv}
\partial_t{\mathcal S} = 
-{\rm Tr}({\rm ln} \hat \rho \, \partial_t \hat \rho ) = 
\tfrac{g_{\mu\nu}}{\hbar} \, {\rm Re}\,
\langle \hat L_\mu^\dag [\hat L_\nu, {\rm ln} \hat \rho]\rangle,         
\end{equation}
which is also a state function. 
For concreteness,    
this conclusion can be achieved by inserting 
(\ref{eq:GaugeTrans}) into (\ref{eq:entEv}).  
The similarity of the above equation with the current in Eq.(\ref{eq:powcurr})
is not a coincidence, it actually refers to 
the statistical interpretation \cite{balian2007}:  
currents are bonded to entropy changes 
since heat is associated with the variation of populations 
$d\hat \rho  = \partial_t \hat \rho \, dt$, 
see Sec.\ref{sec:SIHW}. 

The contraction of the relative entropy 
$S(\Phi_t[\hat \rho_0]|\Phi_t[\hat \rho_\star]) $ 
$\leq S(\hat \rho_0|\hat \rho_\star)$ \cite{lindblad1975} 
under the \hy{CPTP} map $\Phi_t$ in (\ref{eq:dynmap}) 
associated with the \hy{ME} in (\ref{eq:mastereq}) 
enables us to define, as in the strictly Markovian case \cite{spohn1978,BreuerPetr2002}, 
the entropy production (\hypertarget{EP}{EP})
\begin{equation}\label{eq:entProd}
\Sigma := S(\hat \rho_0|\hat \rho_\star) - S(\hat \rho_t|\hat \rho_\star) \geq 0
\end{equation}
and its rate 
\begin{gather}\label{eq:entProdRate}
{\mathcal E} := \partial_t\Sigma = \partial_t\mathcal{S} + 
\partial_t {\rm Tr} ( \hat \rho_t \, {\rm ln} \hat \rho_\star ), 
\end{gather}
where $\partial_t {\mathcal S}$ is in \eqref{eq:entEv}, 
while the remaining term is related to the entropy flux 
due to heat exchange between $\text{S}$ and $\text{E}$.
In the above formulas, $\hat \rho_\star$ is a fixed point of the \hy{CPTP} map, 
\emph{i.e.}, a solution of $\Phi_t[\hat \rho_\star] = \hat \rho_\star$. 

Since $\mathcal S$ and $\partial_t {\mathcal S}$ are state functions, 
one can easily gather that both $\Sigma$ and ${\mathcal E}$ 
are gauge-invariant, or also state functions.
For a non-Markovian evolution, 
the \hy{EP} rate in Eq.\eqref{eq:entProdRate} can be momentarily negative, 
a feature associated with the break of P-divisibility of $\Phi_t$ \cite{breuer2016}, 
while $\Sigma$ is always non-negative.
The (non)positivity of the \hy{EP} rate, as stated in this paragraph, 
is a vocable for the Second Law of thermodynamics in the quantum realm,  
and its violation (break of P-divisibility) is associated with 
non-Markovian effects \cite{breuer2016}. 

Another formula for an \hy{EP} rate found in \cite{colla2022} 
considers a situation in which system $\rm S$ is strongly
coupled with a thermal bath (system $\rm E$) at inverse temperature $\beta$: 
\begin{equation}\label{eq:entProdRate:th}
\tilde{\mathcal E} = 
\partial_t{\mathcal S} - \beta \mathcal J 
%
= \tfrac{g_{\mu\nu}}{\hbar} \, {\rm Re}\,
\langle \hat L_\mu^\dag [\hat L_\nu, \beta \hat H + 
{\rm ln} \hat \rho]\rangle,         
\end{equation}
where 
$\beta \mathcal J$ 
is the flux associated with the total current in (\ref{eq:powcurr}).   
Despite positive for a P-divisible map 
(equivalent to a Markovian \hy{ME})
$\Phi_t$ \cite{colla2022,breuer2016},  
the expression for 
$\tilde{\mathcal E}$ will be invariant 
if and only if conditions (\ref{eq:invCond}) 
are satisfied, due to its dependence on 
$\mathcal J$. 
On another side, 
choosing a gauge like $(iii)$ in Sec.\ref{sec:AnParProc} 
ensures the preservation of $\tilde{\mathcal E}$.
Interestingly enough, 
$\tilde{\mathcal E}$ changes according to the current, 
see Eq.(\ref{eq:gCur}), 
\begin{equation}\label{eq:gEntProd}
\tilde{\mathcal E}' = \tilde{\mathcal E} 
- \beta ({\mathcal J}' - {\mathcal J}) = 
\tilde{\mathcal E} - \beta ( {\mathcal J}_{\! \delta \! \hat H} +
      {\mathcal C}_{\! \delta \! \hat H} ), 
\end{equation}
in order to keep the invariance of $\partial_t\mathcal S$.  

The expression in \eqref{eq:entProdRate:th}
was deduced from a specific gauge choice in \cite{colla2022}, 
the minimal dissipation gauge, item $(iv)$ in Sec.\ref{sec:AnParProc},  
therefore there is no strangeness on its gauge dependence.
The authors there consider an instantaneous Gibbs 
state of the Hamiltonian $\hat H$ present in the \hy{ME}, 
the gauge-transformed version in (\ref{eq:gEntProd}) 
is obtained considering the evolution of the system
governed by the transformed Hamiltonian 
$\hat H + \delta \! \hat H$. 

The universal formulation of the \hy{EP}, {\it i.e.}, 
for arbitrary subsystems ($\rm E$ and $\rm S$) 
and arbitrary dynamics, 
considers the system, the environment, 
and their joint unitary evolution \cite{esposito2010}. 
Part of this information is suppressed by the trace in (\ref{eq:dynmap}) 
and the expressions $\mathcal E$ and $\tilde{\mathcal E}$ 
were developed for specific \hy{ME} scenarios. 
While Eq.(\ref{eq:entProdRate}) 
simply raises from the contraction of $\Phi_t$, 
it requires explicitly a fixed point $\hat \rho_\star$, 
which is not a generic property of \hy{ME}s. 
On another side, assuming that system $\rm E$ is a thermal environment, 
Eq.(\ref{eq:entProdRate:th}) deduced for the \hy{ME} \cite{colla2022} 
coincides with the thermodynamical \hy{EP} in \cite{deffner2011}.

\section{Example: Pure Decoherence Model}\label{sec:ExPDM}
In the wake of open quantum systems, decoherence has a special appeal 
due to its relation with the quantum-classical border \cite{Schlosshauer2008}. 
In this section, 
we will study the gauge-thermodynamical behavior 
of the simpler non-Markovian system 
that captures the essence of this phenomenon, 
the pure decoherence model \cite{breuer2016,BreuerPetr2002}.

Let us consider the \hy{ME} (\ref{eq:mastereq}) for a two-level system 
with only one Lindblad operator and a Hamiltonian, respectively, given by 
\begin{equation}\label{PDM:hamlind}
\hat L = \sqrt{\hbar|\Gamma(t)|} \, \hat\sigma_z,  \,\,\, 
{\hat H} = \tfrac{\hbar}{2} \omega \hat\sigma_z,   
\end{equation}
where 
$\hat\sigma_z$ is the standard Pauli matrix, 
$\omega$ is the transition frequency,
and 
$\Gamma(t) \in \mathbb{R}$ is the decay rate, 
in principle, a generic time-dependent function.  
For this system, Eq.(\ref{eq:mastereq}) has 
$M = 1$ and ${\bf g} =:g = \text{sign}[\Gamma(t)]$.  

The solution for the \hy{ME} is easily obtained,  
\begin{gather}\label{PDM:state} 
\rho_{00} = p = 1 - \rho_{11}, \,\,\, 
\rho_{01}(t) = \rho_{01} \, \text{e}^{-i\omega t} \, D(t), \,\,\, 
\end{gather}
where $\rho_{ij}(t) = \bra{i} \hat \rho(t) \ket{j}$ 
are the matrix elements of the density operator and 
we denote $\rho_{ij}= \rho_{ij}(t=0)$. 
The initial populations 
$0 \leq p \leq 1$ and $1-p$ are constant, 
which is not true for the coherence term 
$\rho_{01} = \rho_{10}^\ast \in \mathbb{C}$, 
the latter evolves according to the decoherence 
function $D(t) := \exp\{-\int_0^t \Gamma(s)ds\}$, 
which encapsulates all the decoherence properties 
of the system \cite{BreuerPetr2002,Schlosshauer2008}.  

Directly from the solution above,
for a given initial state with elements $\rho_{00} = p$ and $\rho_{01}$, 
the state 
\begin{equation}\label{PDM:fixPt}
\hat \rho_\star = p \ket{0}\bra{0} + (1-p) \ket{1}\bra{1}
\end{equation}
is a fixed point of the dynamics. 
The long-term solution of the \hy{ME} will converge as 
$\hat \rho (t\to \infty) = \hat \rho_\star$, 
if the decoherence effects destroy completely 
the initial-state coherence $\rho_{01}$, 
which will happen for $\lim_{t \to \infty} D(t) = 0$.
The evolution will be Markovian if $D(t)$ decreases monotonically 
since the environment will progressively erasure the coherences.
Otherwise, a non-monotonic behavior of $D(t)$ will raise non-Markovian effects
since the environment will create coherence 
in the system \cite{BreuerPetr2002,breuer2016,Schlosshauer2008}.  
Recalling that, to represent a physical system, 
$\hat \rho$ must be a positive operator, 
which implies $|\rho_{01}(t)| \le (p-p^2), \forall t$,
thus $|D(t)| \le 1$. 
At least theoretically, 
one can consider any function $D(t)$ satisfying the positivity restriction, 
nevertheless, the system may not converge to $\hat\rho_\star$,  
which will still be a fixed point of the \hy{ME}. 

For the thermodynamical analysis, we first consider the path $\ell_0$ 
defined in (\ref{eq:pathl0}) for the operators in (\ref{PDM:hamlind}). 
The current in (\ref{eq:sumcurr}) is null, 
${\mathcal J} = - \tfrac{g}{\hbar} \, {\rm Re} \,
\langle \hat{L}_\mu^\dagger [ \hat{L}_\nu , \hat{H} ]  \rangle = 0$,
once $[\hat H,\hat L] = 0$. 
The power in \eqref{eq:powcurr} is also null, 
${\mathcal P} = 0$, since $\partial_t \hat H = 0$. 
Consequently, 
the energy $\langle \hat H \rangle = 
\tfrac{1}{2}\hbar\omega (2p-1)$ is constant for any initial state, 
according to (\ref{eq:firstlaw}).  
This clearly agrees with (\ref{eq:heatworkc0}), 
which says that there is no heat exchange and no work performed 
by/on the system in path $\ell_0$. 

Using the expressions in (\ref{PDM:state}), we find
\[
\langle \hat L \rangle = (2p-1)\, \sqrt{|\Gamma(t)|}, \,\,\, 
\langle \partial_t \hat L \rangle = (2p-1)\, \partial_t \sqrt{|\Gamma(t)|}. 
\] 
Hence, according to conditions (\ref{eq:invCond}), 
non-equilibrium thermodynamical invariance is fulfilled solely 
for states with equally populated levels, {\it i.e.}, 
$p = 1/2$. 
Since $\rho_{00}$ and $\rho_{11}$ are constant throughout the whole evolution,
if the initial state is equally populated, 
the system will be thermodynamically invariant for any time.   
For such an initial state, 
the values of the thermodynamical quantities in $\ell_0$, 
$\partial_t\langle \hat H \rangle = 
{\mathcal J} = {\mathcal P} = 0$,  
are the same for any gauge transformation (\ref{eq:GaugeTrans}) 
with $\partial_t \phi = 0$, 
which for this system is written as 
\[
\hat L' = {\rm e}^{i\theta(t)}\hat L + \gamma(t), 
\,\,\, 
\delta\!\hat H = g \sqrt{|\Gamma(t)|} 
\text{Im}[\gamma(t)] \hat \sigma_z + \phi. 
\]

Each of the above gauge transformations defines a path $\ell$ 
according to (\ref{eq:pathl}), 
for simplicity in what follows we will consider paths such that 
$\theta(t) = 0$ and that $\gamma(0) = 0$. 

For a non-invariant initial state, 
the internal energy in $\ell$ is 
\[
\begin{aligned}
\langle \hat H' \rangle &= 
\langle \hat H\rangle + \langle\delta\!\hat H \rangle \\
&= 
\left(\tfrac{1}{2}\hbar\omega + g \sqrt{|\Gamma(t)|} 
\text{Im}[\gamma(t)] \right) (2p-1) + \phi,     
\end{aligned}
\]
which is no more constant, $\partial_t\langle \hat H' \rangle \ne 0$.
Heat current also vanishes in $\ell$,
$\mathcal{J}'=\mathcal{J} = 0$, because 
$\mathcal{J}_{\delta\!\hat H} = \mathcal{C}_{\delta\!\hat H} = 0$ 
in Eq.\eqref{eq:curpowdeltaH}; 
therefore within this generic path, 
the work is equal to the change in internal energy:  
\begin{gather}\label{pdm:work}
W_\ell = \int_{0}^{t}\partial_t \langle \delta\!\hat H \rangle dt = 
(2p-1) g \, \sqrt{|\Gamma(t)|} \, \text{Im}[\gamma(t)]\, ,
\end{gather}
see Eq.(\ref{eq:heatworkc}). 
Its rate provides the power 
${\mathcal P'} = \partial_t \langle \delta\!\hat H \rangle$ along $\ell$. 

The \hy{EP} in (\ref{eq:entProd}) is a state function, 
meaning that it has the same value for any path $\ell$, 
which is equal to its value in $\ell_0$. 
Diagonalizing the operator $\hat \rho(t)$ in \eqref{PDM:state}, 
its eigenvalues are $\tfrac{1}{2} \pm R(t)$, where 
\[
R(t) := \sqrt{(p-\tfrac{1}{2})^2 + |\rho_{01} D(t)|^2}\le 1/2, 
\forall t \ge 0,
\]
is a real bounded function 
as a consequence of the positivity of $\hat \rho(t)$;  
note that $D(0) = 1$ and $|D(t)| \le 1$.
Using (\ref{PDM:fixPt}) and the eigenvalues of $\hat \rho(t)$, 
it is possible to write the \hy{EP} as 
\[
\begin{aligned}
&\Sigma  = \text{H}[R(t)] - \text{H}[R(0)] \ge 0, \\
&\text{H}(x) := - (\tfrac{1}{2}+x)\ln(\tfrac{1}{2}+x)
               - (\tfrac{1}{2}-x)\ln(\tfrac{1}{2}-x).
\end{aligned}
\]
The function $\text{H}$ 
is Schur-concave \cite{Marshall2010} for $0 \le x \le \tfrac{1}{2}$, 
which implies $\text{H}[R(t)] \ge \text{H}[R(0)]$, 
thus $\Sigma \ge 0$, as expected.

The \hy{EP} rate becomes
\[
{\mathcal E} = -\partial_t\text{H}[R(t)] 
= - \frac{|\rho_{01}|^2}{R(t)} \text{Arctgh}[2R(t)] \, 
\partial_t[D(t)]^2, 
\]
which is positive whenever $\partial_t D < 0$.  
Recalling that a monotonically decreasing $D(t)$ 
corresponds to a Markovian dynamics,  
non-Markovianity emerges for functions 
$D(t)$ with $\partial_t D > 0$, 
which will also give rise to a negative \hy{EP} rate, 
and is associated with violations 
of the Second Law for quantum systems \cite{breuer2016, colla2022}.

The thermodynamical behavior of decoherence is similar to 
the classical experiment of ``free expansion'' \cite{callen1985},  
where a gas irreversibly expands 
without exchanging heat or performing work, 
due exclusively to positive entropy production. 
This is the same behavior of the example for 
an invariant initial state, 
or even for the system following the path $\ell_0$, 
where there are no energetic changes, 
only entropic ones associated with $\Sigma \ge 0$. 
Further, 
for the same classical experiment, 
when isolated and expanding against a piston, 
there will be entropy production and work will be realized by  
the gas. This situation is comparable to any path other than $\ell_0$ 
followed by the quantum system, 
where the power is not null and provides work in (\ref{pdm:work}). 
Noteworthy, this analogy is lawful even when non-Markovianity 
takes place in the dynamics of the quantum system 
since $\Sigma \ge 0$ regardless of the sign of $\mathcal E$.
%
\section{Application to Thermal Machines}\label{sec:atm}
A quantum thermal machine 
is nothing but a system evolving cyclically with period $\tau$, 
while interacting with reservoirs.  
The whole system evolution, 
a closed ther\-mo\-dy\-namical path $\ell_0$, 
see Eq.(\ref{eq:pathl0}), 
is described by periodic operators, {\it i.e.},
$\hat H(t + \tau) = \hat H(\tau)$,  
$\hat L_\mu(t + \tau) = \hat L_\mu(t), \forall \mu$.   
For this path, heat and work in one period
are given by (\ref{eq:heatworkc0}), 
which in this case are closed integrals,
\begin{equation}\label{eq:macHW}
\begin{aligned}
Q_{\ell_0} &= \oint_{\ell_0} {\mathcal J} dt,
\,\,\,\,
{W}_{\mathcal \ell_0} = \oint_{\ell_0} {\mathcal P} dt,
\end{aligned}
\end{equation}
for $\mathcal J$ and $\mathcal P$ in \eqref{eq:powcurr}
written for the Hamiltonian and Lindblad operators
governing the system evolution.

For a machine composed of two thermal reservoirs, 
a colder $\rm c$ and a hotter $\rm h$, 
as in the classical Carnot system \cite{callen1985},
the total current in (\ref{eq:powcurr}) 
will be the sum (\ref{eq:sumcurr}).
However, each reservoir may be
described by more than one Lindblad operator,
which leads us to define 
$\mathcal{J}_\text{c}$ and $\mathcal{J}_\text{h}$
as sums of the currents of the respective Lindblad operators
and write the total current as
$\mathcal{J} = \mathcal{J}_\text{c} + \mathcal{J}_\text{h}$.
Consequently, 
the heat entering the system from the hotter reservoir 
and the heat leaving the system to the colder are, respectively,
written as
\begin{equation} \label{eq:macHeats}
Q^\text{h}_0 := \oint_{\ell_0} \mathcal{J}_\text{h} \, dt ,
\,\,\,\,
Q^\text{c}_0 := \oint_{\ell_0} \mathcal{J}_\text{c} \, dt,
\end{equation}
and, according to (\ref{eq:macHW}),
$Q^\text{c}_0 + Q^\text{h}_0 = Q_{\ell_0}$.

As the system returns to the initial state,
the mean energy assumes its initial value,
since the Hamiltonian is periodic.
Consequently,
$W_{\ell_0} + Q_{\ell_0} = 0$
is the integral expression of the First Law (\ref{eq:firstlaw})
for a whole cycle of the machine.
The efficiency of the machine, as in a classical cycle,
will be defined by the ratio
\begin{equation} \label{eq:macEffic}
\eta = |W_{\ell_0}| / Q^\text{h}_0 =
1 -  |Q^\text{c}_0| / Q^\text{h}_0.
\end{equation}

The von Neumann entropy also returns to its initial value after a cycle
due to the periodicity of the Lindblad operators, see (\ref{eq:entEv}).
Using Eq.(\ref{eq:entProdRate:th}),
\begin{equation*}
\Delta {\mathcal S} =
\oint_{\ell_0} (\tilde{\mathcal E} +
\beta{\mathcal J})dt = 0.
\end{equation*}
Defining the \hy{EP} as
$\tilde \Sigma := \oint_{\ell_0} \tilde{\mathcal E} dt$,
using \eqref{eq:macHW} and (\ref{eq:macHeats}), and performing the integration,
one finds
\begin{equation}\label{eq:macEntProd}
\tilde \Sigma +
\beta_\text{h} Q^\text{h}_0 +
\beta_\text{c} Q^\text{c}_0 = 0.
\end{equation}
The positivity of the \hy{EP}, $\tilde \Sigma \ge 0$, 
implies $\beta_\text{c} |Q_\text{c}^0| \ge \beta_\text{h} Q_\text{h}^0$,
which replaced in (\ref{eq:macEffic})
shows that the efficiency is bounded by the Carnot limit, {\it i.e.},
\begin{equation}\label{eq:eff}
\eta =
1 -  |Q^\text{c}_0| / Q^\text{h}_0 
\le 1 - \beta_\text{h}/\beta_\text{c}.
\end{equation}

To take into account the gauge effects, 
we choose a closed thermodynamical path $\ell$, 
see Eq.(\ref{eq:pathl}), 
described by periodic gauge-functions 
$\{\gamma_\mu, U_{\!\mu\nu}, \phi\}$.  
In this way, the Hamiltonian  
$\hat H' = \hat H + \delta\!\hat H$ 
and the Lindblad operators $\hat L'_\mu$
will be also periodic operators.
Non-periodic gauge functions (open paths $\ell$), 
could be equally considered, 
though thermodynamical quantities 
will not return to their initial value after one period, 
a necessary requisite for a proper machine.  

The heat $Q_{\ell}$ and work $W_{\ell}$
will be given by (\ref{eq:heatworkc}) and 
formulas (\ref{eq:heatworkVar}) remain valid for the closed integral.
The current after a gauge (\ref{eq:gCur}) 
is the sum of the currents of each reservoir, 
${\mathcal J}' = {\mathcal J}'_\text{c} + {\mathcal J}'_\text{h}$, 
as well $Q_\text{c}^\ell + Q^\ell_\text{h} = Q_{\ell}$ will be 
total heat with 
\begin{equation}\label{eq:macGHheats}
Q^\text{h}_\ell := \oint_{\ell} \mathcal{J}'_\text{h} \, dt ,
\,\,\,\,
Q^\text{c}_\ell := \oint_{\ell} \mathcal{J}'_\text{c} \, dt.
\end{equation}

As before, after a period $W_{\ell} + Q_{\ell} = 0$, 
although $W_{\ell_0} + Q_{\ell_0} = 0$.
Consequently,
using (\ref{eq:heatworkVar}) and (\ref{eq:gFirstLaw}),
\begin{equation*}
\begin{aligned}
\oint \partial_t \langle \delta\! \hat H \rangle dt = 0
\end{aligned}
\end{equation*}
is in agreement with the periodicity of $\delta\! \hat H$ and 
is independent of the path $\ell$.

For the \hy{EP}, 
the periodicity of the von Neumann entropy gives
\begin{equation} \label{eq:gMacEntChange}
\oint_{\ell} (\tilde{\mathcal E}' +
\beta{\mathcal J'})dt =  
\tilde \Sigma' +
\beta_\text{h} Q^\text{h}_\ell +
\beta_\text{c} Q^\text{c}_\ell = 0, 
\end{equation} 
which, from the positivity of the \hy{EP} 
$\tilde \Sigma' := \oint_{\ell} \tilde{\mathcal E}' dt \ge 0$, 
ensures that the efficiency, despite not gauge-invariant, 
is also bounded by the Carnot limit for any gauge: 
\begin{equation}\label{eq:gMacEffic}
\eta' = |W_{\ell}|/{Q}_{\ell}^\text{h} =
1 -  |{Q}_{\ell}^\text{c}|/{Q}_{\ell}^\text{h} \le 
1 - \beta_\text{h}/\beta_\text{c}. 
\end{equation}
%

\subsection{General Comments}
Despite the lack of a general theory for the \hy{EP} for \hy{ME}, 
see Sec.\ref{sec:EntProd},  
the obtainment of \eqref{eq:macEntProd} or \eqref{eq:gMacEntChange} 
was only possible due to the recent developments in \cite{colla2022}, 
which derives the \hy{EP} for a system strongly coupled to a thermal reservoir.
In some sense, excepting the gauges,  
the machine description and treatment is commonly found  
in the literature on Markovian machines, 
see \cite{alicki2018,scopa2018} and the references therein. 

While Eq.(\ref{eq:eff}) refers to a machine 
where the dynamics is ruled by a \hy{ME} (\ref{eq:mastereq}),
but describing the evolution of a system (strongly) coupled to thermal reservoirs, 
see \cite{colla2022}, 
its particular Markovian version \cite{Note:AlickiDef} is identical and 
was derived in \cite{alicki1979}, 
including the limitation by the Carnot Bound. 
The efficiency and the Carnot bound were studied for 
quantum machines which do not fit in both of these scenarios,  
for instance, 
generic Markovian master equations with fixed points \cite{gardas2015}, 
the replacement of a thermal bath by a squeezed version \cite{niedenzu2018},
and also a machine governed by a global unitary evolution \cite{skrzypczyk2014}. 

The formulas for the \hy{EP}, 
Eqs.\eqref{eq:macEntProd} or \eqref{eq:gMacEntChange}, 
are promptly obtained if the closed path is composed, 
as in a Carnot machine, by strokes: 
an isothermal at a definite temperature, 
representing the coupling with a reservoir, 
followed by an isentropic, in which the evolution is unitary.
An evolution like this may be devised using periodic Lindblad operators 
becoming cyclically null throughout the isentropics. 
The \hy{EP} in each isothermal 
will be given by Eq.(\ref{eq:entProdRate:th}), 
while it will be null throughout the isentropic 
since the unitary evolution ensures 
${\mathcal J} = 0$ and $\partial {\mathcal S}_t = 0$.    
For the whole cycle, 
the \hy{EP} will be (\ref{eq:macEntProd}) with 
$Q_0^\text{h}$ and $Q_0^\text{c}$, the heat exchanged 
along the hot and the cold isotherms, respectively. 
The positivity of $\tilde\Sigma$ in each isotherm, 
given by (\ref{eq:entProdRate:th}), 
is ensured if the evolution is Markovian \cite{colla2022}, 
which gives the limitation by the Carnot bound. 

Reciprocating machines, 
the ones in which combined stro\-kes describes the evolution,
are suitable for analytical calculations     
due to the convenient separation between 
the generators of the \hy{ME} \cite{alicki2018}, 
as described by the Carnot-like machine above. 
``Continuous-time'' quantum machines \cite{alicki2018} 
are the ones in which the path can not be separated in strokes.  
In this case, 
the integral of $(\tilde{\mathcal E} + \beta{\mathcal J})$ 
can be split as a sum of many infinitesimal connected paths, 
in which each alternate path represents the evolution 
of the system interacting with only one reservoir,  
and Eq.(\ref{eq:entProdRate:th}) 
can be applied in each piece.
This scenario can be theoretically described 
in the scope of a Floquet theory for \hy{ME}s \cite{scopa2018}. 

The positivity of the \hy{EP}, a statement of the Second Law, 
ensures the limitation of the efficiency by the Carnot bound 
in (\ref{eq:macEffic}) and in (\ref{eq:gMacEffic}).  
However, it is also a necessary condition for Markovian dynamics, 
see Sec.\ref{sec:EntProd}, 
thus a violation of the Second Law may happen for non-Markovian 
evolutions and that bound may be surpassed.  
This would happen for any gauge 
since the dynamics is invariant under gauge transformations. 

\subsection{Gauge and Efficiency}
Our previous results show that all gauges of a thermal
machine described by a Markovian \hy{ME} are subjected 
to the Carnot bound,  
however, it remains to show how the efficiency 
changes according to a gauge choice. 
We will tackle this question for 
a thermodynamical-invariant thermal machine, 
which unexpectedly has its efficiency changed, 
despite the invariance.  

The main point is to take into consideration that 
the invariance conditions in (\ref{eq:invCond})
do not apply for each current $\mathcal J_\mu$ in (\ref{eq:gLindCurr})
but only for the whole $\mathcal J$ in (\ref{eq:gCur}).
In other words, ${\mathcal J}'_\mu \ne {\mathcal J}_\mu$ in general, 
even for an invariant system where $\mathcal J' = \mathcal J$.  
Consequently, in general,
$\mathcal{J}'_\text{c} \ne \mathcal{J}_\text{c}$ and 
$\mathcal{J}'_\text{h} \ne \mathcal{J}_\text{h}$, 
as the sum of the Lindblad operators relative 
to their respective reservoirs.  
However, the invariance of $\mathcal J$ and 
definitions Eqs. \eqref{eq:macHeats} and \eqref{eq:macGHheats} 
ensure 
\begin{equation}\label{eq:gCurRel}
{\mathcal J}'_\text{c} + {\mathcal J}'_\text{h} = 
{\mathcal J}_\text{c} + {\mathcal J}_\text{h}
\Longleftrightarrow 
(Q^\text{h}_\ell - Q^\text{h}_0) = -
(Q^\text{c}_\ell - Q^\text{c}_0).  
\end{equation}

Using (\ref{eq:macEntProd}),
one can rewrite Eq.(\ref{eq:gMacEntChange}) as
\begin{equation}\label{eq:macVarEP}
\begin{aligned}
\tilde \Sigma' - \tilde \Sigma &= 
- \beta_\text{h} (Q^\text{h}_\ell - Q^\text{h}_0) 
- \beta_\text{c} (Q^\text{c}_\ell - Q^\text{c}_0) \\
&= 
(\beta_\text{c}-\beta_\text{h}) (Q^\text{h}_\ell - Q^\text{h}_0),  
\end{aligned}
\end{equation}
which is the integral formulation of Eq.(\ref{eq:gEntProd});
the second equality is obtained using the relation (\ref{eq:gCurRel}). 
The last equality in (\ref{eq:macVarEP}) 
is symmetric concerning the paths $\ell$ and $\ell_0$,
reflecting that both paths are equally possible gauges of the same evolution.  
It asserts that 
if the entropy production (considered positive) is bigger in one gauge, 
the positive heat will be greater than in the other.   
The machine efficiency will change accordingly 
due to the thermodynamical invariance. 
Since $\mathcal P = \mathcal P'$,  
thus $|W_{\ell}| = |W_{\ell_0}|$, 
which gives 
$\eta' = |W_{\ell}|/{Q}_{\ell}^\text{h} = 
|W_{\ell_0}|/{Q}_{\ell}^\text{h}$,
and the machine efficiency increases 
for a reduction of the entropy production 
and {\it vice-versa}. 

For a non-invariant system, 
the work performed and heat absorbed both 
changes with gauges, 
and generically nothing more can be said besides the limitation 
of the efficiency by the Carnot bound, 
as stated in (\ref{eq:gMacEffic}).

\subsection{Example: The Three-Level Maser}
\begin{figure}[htbp]
\includegraphics[width=0.8\columnwidth]{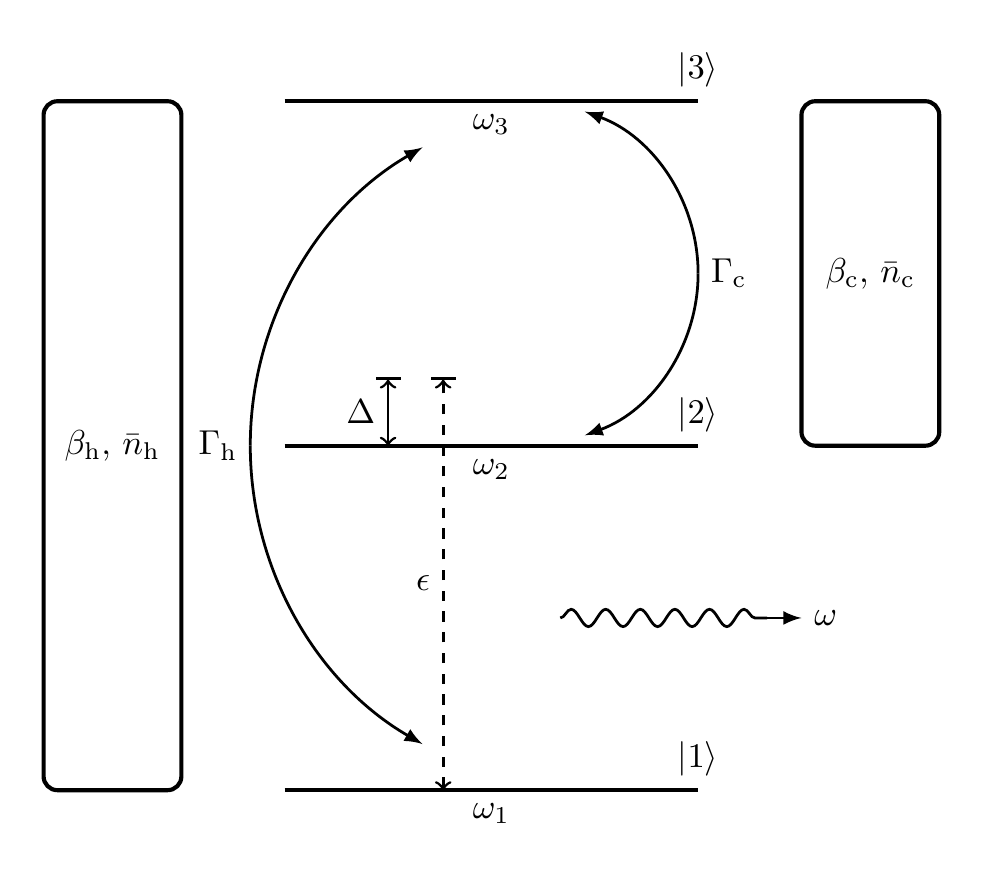}
\caption{Scheme of a three-level quantum 
thermal machine for a system with energies 
$\hbar \omega_m$ for $m = 1,2,3$. 
Bath $\alpha$ with population 
${\bar n}_\alpha = {\bar n}_\alpha := 
( \text{e}^{\beta_\alpha \hbar \omega_\alpha } - 1 )^{-1}$ 
and inverse temperature $\beta_\alpha$ 
induces transitions with rate $\Gamma_\alpha$ 
for $\alpha \in \{\text{c},\text{h}\}$
with 
$\omega_\text{h} = \omega_3-\omega_1$ and 
$\omega_\text{c} = \omega_3-\omega_2$. 
Levels $1$ and $2$ are connected by 
an external classical field $\hat V(t)$
and, at the end of the cycle, 
a photon is emitted with frequency 
$\omega$. 
This figure is largely inspired 
by Fig.1 of \cite{kalaee2021}.}              \label{fig:qtm}
\end{figure}

The quantum thermal engine introduced in \cite{scovil1959} 
considers a three-level atom coupled to two thermal baths 
and to a radiation field, as depicted in Fig.\ref{fig:qtm}. 

The Hamiltonian of this system is 
$\hat H = \hat H_0 + \hat V(t)$ as in \cite{kalaee2021}, 
with 
\begin{equation*}
\begin{aligned}
\hat H_0 &= \sum_{m=1}^3 \hbar \omega_m \, \ket{m}\bra{m} \, , \\
\hat V(t) &= \hbar \epsilon 
( \text{e}^{i\omega t} \ket{1}\bra{2} + 
  \text{e}^{-i\omega t} \ket{2}\bra{1} ) \, ,
\end{aligned}
\end{equation*}
where $\hat H_0$ governs the dynamics of the atomic levels 
with $\omega_1 < \omega_2 < \omega_3$, 
and the atom-field interaction potential $\hat V(t)$ 
has coupling constant $\epsilon$, 
while $\omega$ is the field angular frequency. 

Each thermal reservoir 
is described by two Lindblad operators, 
corresponding to thermal transitions between the energy levels: 
\begin{gather*}
\hat A_{\text{h},1} = \sqrt{\hbar\Gamma_\text{h} {\bar n}_\text{h}}\, \ket{3}\bra{1}
\, ,\quad 
\hat A_{\text{h},2} = \sqrt{\hbar\Gamma_\text{h} ({\bar n}_\text{h}+1)}\, \ket{1}\bra{3}
\, ,
\end{gather*}
for the hotter at ``inverse temperature'' $\beta_\text{h}$, while
\begin{gather*}
\hat A_{\text{c},1} = \sqrt{\hbar\Gamma_\text{c} {\bar n}_\text{c}}\, \ket{3}\bra{2}
\, ,\quad 
\hat A_{\text{c},2} = \sqrt{\hbar\Gamma_\text{c} ({\bar n}_\text{c}+1)}\, \ket{2}\bra{3}
\, ,
\end{gather*}
are associated with the colder at temperature $\beta_\text{c}$. 

The \hy{ME} governing the system evolution is written as (\ref{eq:mastereq})
with $M = 4$, $\mu\in \{(\text{h},1),(\text{h},2),(\text{c},1),(\text{c},2)\}$, 
and $g_{\mu\nu} = \delta_{\mu\nu}$. 
From now on and without loss of generality, 
we will use $\Gamma_\text{h} = \Gamma_\text{c} =: \Gamma$ 
and $\Delta =0$. 

After a transient, 
the system attains a limit cycle, 
the path $\ell_0$ in (\ref{eq:pathl0}), 
described by a periodic density matrix 
given by
\begin{equation*}
\hat \rho(t) = \hat \rho(t+\tau)=
\left( \begin{array}{ccc}
\rho_{11} & \rho_{12} {\rm e}^{i\omega t} & 0 \\
\rho_{12}^\ast {\rm e}^{-i\omega t} & \rho_{22} & 0 \\
0 & 0 & \rho_{33}
\end{array} \right),
\end{equation*}
where the period of the cycle is determined by 
$\omega = \omega_2-\omega_1 = 2\pi/\tau$ and 
the elements $\rho_{ij}$ are constants  
($\rho_{jj} \in \mathbb R$ and $\rho_{12} \in \mathbb C$),  
given by \cite{kalaee2021} 
\begin{equation*}
\begin{aligned}
\rho_{11} &= [\Gamma {\bar n}_\text{c}
   ({\bar n}_\text{h}+1) + \lambda ({\bar n}_\text{c}+{\bar n}_\text{h}+2) ]K^{-1},  \\
\rho_{22} &= [\Gamma ({\bar n}_\text{c} + 1){\bar n}_\text{h} + 
\lambda ({\bar n}_\text{c}+{\bar n}_\text{h}+2)] K^{-1}, \\
\rho_{33} &=  1-\rho_{11}-\rho_{22},  \,\,\, 
\rho_{12} =  -i \epsilon \lambda^{-1} (\rho_{22} - \rho_{11}),
\end{aligned}
\end{equation*}
%
%
%
where
\begin{equation*}
\begin{aligned}
\lambda &:= 
8 \epsilon^2 \Gamma^{-1} ({\bar n}_\text{c} +{\bar n}_\text{h})^{-2} , \\
K &:= \lambda (4 + 3 {\bar n}_\text{c} + 3{\bar n}_\text{h})  + 
\Gamma ({\bar n}_\text{c} + {\bar n}_\text{h} + 
3 {\bar n}_\text{c} {\bar n}_\text{h})  > 0;
\end{aligned}
\end{equation*}
%
despite the lengthy expressions, 
by now it will be helpful to note only that
$\text{Re}\rho_{12} \le 0$ and  
$\text{Im}\rho_{12} \le 0$. 

The invariance condition in Eq.\eqref{eq:invCond} 
is promptly verified for the periodic solution, since 
$\langle \hat A_{\alpha,j} \rangle = 
\text{Tr} [\hat \rho(t) \hat A_{\alpha,j} ] = 0$ for 
$\alpha \in \{\text{h},\text{c}\}$ and $j=1,2$. 
Thus, for a state in the limit cycle, 
$\mathcal J'= \mathcal J$, $\mathcal P' = \mathcal P$ and 
$\langle\hat H'\rangle = \langle\hat H \rangle$. 
So, heat and work are determined by (\ref{eq:heatworkc0}) for any gauge. 
Noteworthy, in the transient, 
the system will be only gauge-invariant 
if the initial state satisfies Eq.\eqref{eq:invCond}.

The currents associated with the baths are 
$\mathcal{J}_\text{h} = \mathcal{J}_{\text{h},1} + \mathcal{J}_{\text{h},2}$ and  
$\mathcal{J}_\text{c} = \mathcal{J}_{\text{c},1} + \mathcal{J}_{\text{c},2}$, 
where each current is given by (\ref{eq:LindCurr}) with 
$\mu\in \{(\text{h},1),(\text{h},2),(\text{c},1),(\text{c},2)\}$.  
Performing the calculations for 
the above Hamiltonian and Lindblad operators, 
\begin{equation}\label{eq:exCurr}
\begin{aligned}
\mathcal{J}_\text{h} 
%
%
%
&= \hbar \Gamma K^{-1} ({\bar n}_\text{h} - {\bar n}_\text{c}) 
   \lambda \omega_\text{h} \ge 0, \\
\mathcal{J}_\text{c} 
&= -\hbar \Gamma  K^{-1} ({\bar n}_\text{h} - {\bar n}_\text{c}) 
    \lambda \omega_\text{c} \le 0. 
\end{aligned}
\end{equation}
%
%
Although $\mathcal J$ is gauge invariant, 
the above currents in general are not, 
as we will see in a while.  
From Eq.\eqref{eq:powcurr} and above Hamiltonian, 
the power becomes ${\mathcal P} = \langle\partial_t \hat V\rangle$ 
and, due to its invariance, gauge-invariant work 
\begin{equation}\label{eq:exwork}
W_{\ell_0} = \int_0^\tau \langle \partial_t \hat V\rangle dt = 
4\pi\hbar\epsilon \, {\text{Im}}(\rho_{12}) \le 0
\end{equation}
is performed by the system, see Eq.(\ref{eq:heatworkc0}),  
while heat is absorbed from the hotter reservoir.

To compare the efficiencies 
(\ref{eq:macEffic}) and (\ref{eq:gMacEffic}), 
we will choose a closed path $\ell$ in (\ref{eq:pathl}) 
described by the gauge functions 
\[
\gamma_{\text{h},1} = \gamma_{\text{c},1}^* = C {\rm e}^{i\omega t/2}, \,\, 
\gamma_{\text{h},2} = \gamma_{\text{c},2} = \phi = 0, \,\, 
U_{\mu\nu} = \delta_{\mu\nu}, 
\]
where $\omega$ is the same frequency of the state evolution.
Explicitly calculating each $\mathcal J'_\mu$ in (\ref{eq:gLindCurr}),  
using $\mathcal J_\mu$ from (\ref{eq:LindCurr}) and above gauge functions,  
the gauge induces the following change for the hot current:
\begin{gather}\label{eq:exdifCur}
\mathcal{J}_\text{h}' - \mathcal{J}_\text{h} = 
- \tfrac{1}{2}\Gamma \sqrt{ {\bar n}_\text{c} 
               {\bar n}_\text{h} } \, 
|C|^2 \, \text{Im}\!\left(\rho_{12}\right) >0, 
\end{gather}
which integrated over one period gives 
\[
Q_\ell^\text{h} - Q_{\ell_0}^\text{h} = \tau (\mathcal{J}_\text{h}' - \mathcal{J}_\text{h}) > 0.  
\]
From the invariance of $\mathcal J$, see Eq.(\ref{eq:gCurRel}),  
$\mathcal{J}_\text{c}' - \mathcal{J}_\text{c} = - (\mathcal{J}_\text{h}' - \mathcal{J}_\text{h})$ and 
$Q_\ell^\text{c} - Q_{\ell_0}^\text{c} = - (Q_\ell^\text{h} - Q_{\ell_0}^\text{h})$. 
Since $Q_\ell^\text{h} > Q_{\ell_0}^\text{h}$ 
and the work in (\ref{eq:exwork}) is invariant, 
from (\ref{eq:gMacEffic}) and (\ref{eq:macEffic}),   
the gauge transformation thus decreases the efficiency, {\it i.e}, 
$\eta' < \eta$, 
while the \hy{EP} increases, 
see (\ref{eq:macVarEP}).
Replacing $\gamma_{\text{c},1}$ by $-\gamma_{\text{c},1}$, 
while keeping all the other gauge functions, 
the relation (\ref{eq:exdifCur}) will become  
$\mathcal{J}_\text{h}' - \mathcal{J}_\text{h} < 0$. 
In this case, the efficiency and the \hy{EP} 
will respectively increase and decrease.

To show explicitly that the efficiency 
is bounded by the Carnot limit, as stated in (\ref{eq:eff}),  
we thus return to the gauge defined by the path $\ell_0$  
and use the currents in (\ref{eq:exCurr}):
\[
\frac{ |\mathcal{J}_\text{c}|}
     { \mathcal{J}_\text{h}  } 
= \frac{\omega_\text{c}}{\omega_\text{h}} = 
\frac{\beta_\text{h}}{\beta_\text{c}} \, 
\text{ln}\!\left( \frac{\bar n_\text{c}^{-1} + 1 }
                     {\bar n_\text{h}^{-1} + 1} \right) 
\ge \frac{\beta_\text{h}}{\beta_\text{c}}, 
\]
where we inverted the expression for the occupation number $\bar n_\alpha$, 
and the inequality is due to $\bar n_\text{h} \ge \bar n_\text{c}$. 
Using the definition in (\ref{eq:macEffic}) with 
$Q^\alpha_0 = \oint_{\ell_0} 
\mathcal{J}_\alpha dt = \tau \mathcal{J}_\alpha$,  
the efficiency reads
\begin{equation*}
\eta = 1 - |\mathcal{J}_\text{c}| / \mathcal{J}_\text{h} 
= 1 -\omega_\text{c}/\omega_\text{h} \le 
1 -\beta_\text{h}/\beta_\text{c},   
\end{equation*}
as we want to show. 
Trivially, since in path $\ell$ the gauge transformation 
reduces the efficiency, 
this will be also bounded by the Carnot limit, 
agreeing with (\ref{eq:gMacEffic}). 
%


As a last comment, 
not all gauge transformations can change the 
\hy{EP} and the efficiency. 
For instance, choosing    
\[
\gamma_{\text{h},1} = \gamma_{\text{c},1}^* = C {\rm e}^{-i\omega t}, \,\, 
\gamma_{\text{h},2} = \gamma_{\text{c},2} = \phi = 0, \,\, 
U_{\mu\nu} = \delta_{\mu\nu}, 
\]
a very similar gauge to the previous one, 
we find
\begin{gather*}
\mathcal{J}_\text{h}' - \mathcal{J}_\text{h} = 
-\tfrac{1}{2}\Gamma{\sqrt{ {\bar n}_\text{c} 
             {\bar n}_\text{h} } } \, 
|C|^2 \, \text{Im}\!\left(\rho_{12} \,  {\rm e}^{3i\omega t}\right), 
\end{gather*}
which integrated over one period gives 
$Q_\ell^\text{h} - Q_{\ell_0}^\text{h} = 0$, 
and thus neither the efficiency nor the \hy{EP} is changed.  

Finally, 
we remark that in $\ell_0$, 
the Lindblad operators $\hat A_{\mu}$ 
with $\mu\in \{(\text{h},1),(\text{h},2),(\text{c},1),(\text{c},2)\}$ 
have $\text{Tr}\hat A_{\mu} = 0, \forall \mu$, 
and thus satisfy the minimal dissipation condition $(iv)$ 
in Sec.\ref{sec:UnD-NIntSys}. 
The new set of Lindblad operators in $\ell$ is such that 
$\text{Tr}\hat A_{\mu} = \gamma_\mu$ for 
$\mu\in \{(\text{h},1),(\text{c},1)\}$ 
and $\text{Tr}\hat A_{\mu} = 0$
$\mu\in \{(\text{h},2),(\text{c},2)\}$, 
according to Eq.(\ref{eq:GaugeTrans}),
which remove the system from minimal dissipation.  
On another side, 
nonthermal energy sources can change the efficiency 
of a quantum machine \cite{nonthermal,niedenzu2018}, 
which is a behind physical reason explaining 
the gauge-induced efficiency change:    
The operators $\hat A_{\mu}$ in $\ell_0$ describes the interaction 
of the system with thermal reservoirs, 
while in $\ell$ they are ``displaced'' by the gauge,  
$\hat A'_{\mu} = \hat A_{\mu} + \gamma_\mu$, 
representing the coupling with a nonthermal energy source.

\section{Final Remarks and Outlooks}\label{sec:conc}
The \hy{ME} in (\ref{eq:mastereq}) and the thermodynamical quantities 
in (\ref{eq:powcurr}) provide a natural generalization 
of the thermodynamical paradigmatic model constituted by 
a system Markovianly weak-coupled to thermal baths \cite{Note:AlickiDef}, 
developed mainly in \cite{alicki1979,spohn1978}.
In any situation, 
\hy{ME} gauges in (\ref{eq:GaugeTrans}) 
are ascribed to information lack owing to the trace in (\ref{eq:dynmap}). 
If instead, one has access to the global system 
$\hat \rho_0 \otimes \hat \rho_\text{E}$ 
and the global Hamiltonian $\hat H_{\rm SE}$, 
then in this case, 
the unitary evolution will ultimately determine the dynamics 
and the energetic exchanges undergone by the system 
without needing a \hy{ME} description or gauges.
Nonetheless, the debate about thermodynamical definitions 
is open even for global unitary evolution 
\cite{elouard2020,esposito2010,skrzypczyk2014,alipour2016}.

The unavoidable gauge influence on thermodynamics 
was already noted in previous works \cite{kosloff,colla2022}.
Probably the first in recognizing the problem, 
Ref.\cite{kosloff} circumvents the question restricting 
the thermodynamic analysis only to the Markovian weak-coupling-limit 
established by Davies \cite{davies1974}, 
which however is still gauge-dependent, 
while the authors in \cite{colla2022} postulate one specific gauge 
for any \hy{ME},  
the minimal dissipation gauge in Sec.\ref{sec:AnParProc}, 
and develop the whole thermodynamical circumscribed to this choice. 

The crucial point of our work is the identification of 
the gauge-contributions ${\mathcal J}_{\! \delta \! \hat H}$ and 
${\mathcal C}_{\! \delta \! \hat H}$ in (\ref{eq:curpowdeltaH}) 
and $\langle \partial_t \delta \! \hat H \rangle$ 
with the standard statistical interpretation of heat and work, 
see Sec.\ref{sec:SIHW}.  
%
%
Thanks to this agreement, even for a non-invariant system, 
the contributions of heat and work from the interaction 
could be broken unequivocally as in Eq.(\ref{eq:heatworkc}). 
Since ${\mathcal C}_{\! \delta \! \hat H}$ has no classical 
interpretation, see Sec.\ref{sec:SIHW}, 
our definition of heat $Q_\ell$ in (\ref{eq:heatworkc}) is 
the amount of energy exchanged due to the system $\text{E}$ 
and is {\it a posteriori} justified in (\ref{eq:gEntProd}) 
as a contribution to entropy production. 
%
%

We stress the need for a process-dependent interpretation 
linked to a gauge of the \hy{ME}. Without an \emph{a priori} 
physical criteria, there is no way to select one specific 
gauge from the system evolution, see Sec.\ref{sec:GaugeTherm}, 
consequently thermodynamical quantities 
for the same system will be ill-defined. 
Fortunately, 
the mean energy of the system, despite gauge dependent, 
is gauge covariant and legitimizes the First Law for each gauge, 
see Eq.(\ref{eq:gFirstLaw}).  
It is important to recall that even without gauges, 
the First Law needs some modifications to accomplish 
an interplay of terms between the generators, 
which in the end is a gauge transformation, see Sec.\ref{sec:TIAG}. 

A complete thermodynamical description of \hy{ME}s
still lacks an \hy{EP} formulation \cite{landi2021}, 
see Sec.\ref{sec:EntProd}, 
which would end up in an expression for the Second Law.
However, the recent results in \cite{colla2022}
already enabled the construction of a quantum 
thermal machine for systems 
strongly coupled to thermal baths, 
as well as its limitation by the Carnot bound for any gauge of the system, 
as we presented in Sec.\ref{sec:atm}. 
In the scope of \hy{ME}s with fixed points, 
expressions for the \hy{EP} and its rate are known, 
see Sec.\ref{sec:EntProd},  
enabling a complete thermodynamical description, 
which also takes into account the gauges. 
This is exemplified in Sec.\ref{sec:ExPDM} 
with the thermodynamical description of 
the decoherence effect.
In both cases, violations of the Second Law 
related to non-Markovianity \cite{breuer2016,vega2017} 
are discussed.  
These violations are independent of gauges since being Markovian or not 
is a property of the dynamics, which is gauge-invariant. 

Although Alicki in \cite{alicki1979} had undertaken thermodynamical 
definitions for the \hy{ME} in (\ref{eq:mastereq}), 
their interpretation is independent of the dynamics itself.
For instance, the First Law (\ref{eq:firstlaw}) is rewritten 
to emphasize the role of coherences \cite{bernardo2020,alipour2016,ahmadi2022}, 
similar to ${\mathcal C}_{\! \delta \! \hat H}$ in (\ref{eq:curpowdeltaH}).
The interchange among pictures of quantum evolution 
(Schrödinger, Heisenberg, and Dirac)
was the main motivation for another modification of the First Law 
by the inclusion of a reminiscent work term \cite{boukobza2006}, 
which is similar to the gauge-induced power term (\ref{eq:gPower}). 
Gauge transformations like (\ref{eq:GaugeTrans}) 
were not yet put forward for any one of the above proposals 
and may provide interesting results. 

Another perspective for Quantum Thermodynamics 
sets aside a dynamical description by a \hy{ME} 
and focuses on thermodynamical functions, 
in analogy to the Gibbs formulation of potentials 
in classical thermodynamics \cite{callen1985}. 
In a typical scenario, 
the system is pushed away from equilibrium by generalized forces,
often stochastic forces, see \cite{ribeiro2016} and its references.  
For instance, in \cite{deffner2011} 
a system initially in equilibrium with a thermal reservoir 
is isolated from it in order to suffer a unitary evolution.  
At the end of this protocol, 
the employment of the statistical interpretation 
determines the \hy{EP} and 
all relevant thermodynamical functions;
see also \cite{HOAppl}, 
for applications to the harmonic oscillator. 
A similar protocol is put forward in \cite{skrzypczyk2014} 
to extract or perform work from an individual quantum system as 
free energy variation.   
On one side, 
these works provide strong definitions for thermodynamical functions, 
which are immune to gauges.  
On another side, the \hy{ME} conjugated to Alicki's formulas 
is not restricted to specific protocols and environments.

Other approaches join \hy{ME}s and thermodynamical functions 
and one effortlessly proves that the energy or entropy quantities 
are gauge-dependent. 
Defining the informational free energy \cite{gardas2015}, 
the authors presented the correlation heat and the system entropy 
which are gauge-dependent, 
despite the gauge-invariance of informational free energy. 
The consecrated thermodynamical relation between entropy, 
internal energy, and informational free energy is gauge-covariant, 
as in (\ref{eq:TransfmeanH}). 
The works \cite{alah2004, niedenzu2018} adopt 
the concepts of passive energy and ergotropy, 
both gauge-dependent energy pieces 
for a gauge-covariant First Law. 
The interpretation for the gauge-induced modifications, 
as in Sec.\ref{sec:SIHW}, and their consequences 
should be investigated. 

In the end, we will comment on wider applications of the basic concept 
behind our approach. 
Those gauge symmetries are properties of the dynamics, 
thus functions of the system operators $\{\hat H, \hat L_\mu\}$ 
can be affected by gauge transformations. 
For instance: \\
-- The efficiency of a quantum thermal machine is related 
to fluctuations of heat and work \cite{landi2021,kalaee2021} and, 
from the perspective of Thermodynamical Uncertainty Relations (TUR), 
a gauge that increases the efficiency of the machine will increase 
these fluctuations. \\
-- Multi-time correlation functions  
are dynamic functions of operators related 
to experimentally accessible quantities \cite{BreuerPetr2002,khan2022}, 
measures of the system energy 
will change depending on a gauge choice. \\ 
-- Leggett-Garg inequalities constitute tests 
for macrorealistic physical theories \cite{legget1985}. 
Violations or not of such inequalities, 
for systems described by a \hy{ME} \cite{LeggtGargApl},
can be affected by gauges if the operators involved in 
the inequalities were associated with functions of $\{\hat H, \hat L_\mu\}$. \\
%

After the submission of this paper, 
the work \cite{celeri2022} appeared in ArXiv. 
The authors there propose definitions of thermodynamical quantities 
based on the invariance of the mean energy of the system. 
Alicki's thermodynamical functions are replaced by 
Haar-averaged quantities over the unitary subgroup composed
by unitary symmetries of the Hamiltonian operator, 
which they called gauge-emergent symmetry.
Our treatment departs from an opposite direction: 
the non-invariance of the mean energy and the consequential effects 
of the gauge transformations inherent to the \hy{ME} scenario. 
However, 
investigating a similar construction for our results 
is a question to be explored.

\bigskip

\emph{Acknowledgments.---} 
\noindent 
The authors would like to acknowledge the comments of A. Candeloro and 
the anonymous referees.  
F.N. is a member of the Brazilian National Institute of Science 
and Technology for Quantum Information [CNPq INCT-IQ (465469/2014-0)]. 


\end{document}